\documentclass{article}
\pdfoutput=1

\usepackage{arxiv}
\usepackage[utf8]{inputenc} 
\usepackage[T1]{fontenc} 
\usepackage{hyperref} 
\hypersetup{
	colorlinks = true,
	linkcolor = black,
	anchorcolor = black,
	citecolor = black,
	filecolor = black,
	urlcolor = blue
}
\usepackage{booktabs} 
\usepackage{amsfonts} 
\usepackage{nicefrac} 
\usepackage{microtype} 
\usepackage{amsmath,bm}
\usepackage{graphicx}
\usepackage[flushleft]{threeparttable}
\usepackage{multirow}
\usepackage{natbib}
\usepackage{makecell}
\usepackage{caption}
\usepackage{float}
\usepackage{enumitem}
\usepackage{subcaption}
\usepackage{setspace}
\usepackage{textcomp}

\newcommand{\figurehere}[1]{\begin{center}%
	=========================\\%
	Insert Figure #1 about here\\%
	=========================\\%
\end{center}}
\newcommand{\tablehere}[1]{\begin{center}%
	=========================\\%
	Insert Table #1 about here\\%
	=========================\\%
\end{center}}

\usepackage{array}
\newcommand{\PreserveBackslash}[1]{\let\temp=\\#1\let\\=\temp}
\newcolumntype{C}[1]{>{\PreserveBackslash\centering}p{#1}}
\newcolumntype{R}[1]{>{\PreserveBackslash\raggedleft}p{#1}}
\newcolumntype{L}[1]{>{\PreserveBackslash\raggedright}p{#1}}

\title{Extending Mixture of Experts Model to Investigate Heterogeneity of Trajectories: When, Where and How to Add Which Covariates}

\author{
Jin Liu \thanks{CONTACT Jin Liu Email: Veronica.Liu0206@gmail.com, \textcircled{c}2021, American Psychological Association. This paper is not the copy of record and may not exactly replicate the final, authoritative version of the article. Please do not copy or cite without authors' permission. The final article will be available, upon publication, via its DOI: \url{10.1037/met0000436}} \\
Biometrics Department \\
Vertex Pharmaceuticals \\
\And
Robert A. Perera\\
Department of Biostatistics\\
Virginia Commonwealth University \\
}

\begin{document}
\maketitle
\begin{abstract}
Researchers are usually interested in examining the impact of covariates when separating heterogeneous samples into latent classes that are more homogeneous. The majority of theoretical and empirical studies with such aims have focused on identifying covariates as predictors of class membership in the structural equation modeling framework. In other words, the covariates only indirectly affect the sample heterogeneity. However, the covariates' influence on between-individual differences can also be direct. This article presents a mixture model that investigates covariates to explain within-cluster and between-cluster heterogeneity simultaneously, known as a mixture-of-experts (MoE) model. This study aims to extend the MoE framework to investigate heterogeneity in nonlinear trajectories: to identify latent classes, covariates as predictors to clusters, and covariates that explain within-cluster differences in change patterns over time. Our simulation studies demonstrate that the proposed model generally estimates the parameters unbiasedly, precisely and exhibits appropriate empirical coverage for a nominal $95\%$ confidence interval. This study also proposes implementing structural equation model forests to shrink the covariate space of the proposed mixture model. We illustrate how to select covariates and construct the proposed model with longitudinal mathematics achievement data. Additionally, we demonstrate that the proposed mixture model can be further extended in the structural equation modeling framework by allowing the covariates that have direct effects to be time-varying.
\end{abstract}


\setcounter{secnumdepth}{3}
\section{Introduction}\label{intro}
\subsection{Motivating Example}
Multiple existing studies have examined the heterogeneity in mathematics development and how the baseline characteristics inform the formation of such latent classes. For instance, \citet{Kohli2015PLGC1} showed that the clusters of mathematics development exist, and \citet{Liu2019BLSGMM} investigated the impacts of covariates, such as socioeconomic status and teacher-reported scores, have on the heterogeneity in the developmental trajectories of mathematics. These studies lead to an interesting but challenging question: whether these baseline characteristics only affect between-individual differences in within-individual changes in an indirect way. Conceptually, it is reasonable to assume that the covariates that reflect students' ability and potential, such as the teacher-reported approach-to-learning, also directly affect the heterogeneity in mathematics development (i.e., account for the within-cluster variability of trajectories).  

Similar challenges exist in multiple domains. For example, in the biomedical field, cured or uncured latent patient groups may exist for a particular disease. One treatment may affect cured patients more but influence uncured patients less, where the `cured' or `uncured' status may associate with covariates such as demographic information, socioeconomic status, and clinical features. Under such scenarios, researchers usually desire to examine the direct treatment effects on each sub-population and identify covariates that inform patient membership simultaneously, which can be realized by mixture-of-experts (MoE) models. We then extend the MoE to the structural equation modeling (SEM) framework to answer the question presented in the motivating example. 

\subsection{Brief Introduction of Mixture-of-Experts Models}
\citet{Jacobs1991Mixtures} originally proposed the MoE, where the mixing coefficients of mixture components are multinomial functions of covariates to predict clusters; in each component, the outcome variable is a conditional distribution on covariates that account for within-class heterogeneity. Multiple subsequent studies, for example, \citet{Jordan1993MoE}, demonstrated the mixing coefficients can also be other functional forms of covariates. The MoE literature usually terms component densities as `experts' and mixing probabilities as `gating functions'. The notion behind the terminology is that an `expert' can build the conditional distribution in the corresponding covariate space divided by `gating functions' \citep[Chapter~ 14]{Bishop2006pattern}. Essentially, a MoE has three main components: (1) several `experts' that can be any regression functions; (2) `gating' functions that separate the covariate space into several parts with considering uncertainty; more importantly, in each divided region, opinions of the corresponding `expert' are trustworthy; and (3) a probabilistic model that combines gating functions and experts \citep{Jordan1993MoE}. 

Useful statistical models, for example, \citet{Rosen1999MoE, Hurn2003MoE, Carvalho2007MoE, Geweke2007MoE, Handcock2007MoE, LeCao2010MoE} and empirical analyses like \citet{Thompson1998MoE, Gormley2011MoE} with the use of MoEs have been published in multiple areas such as biomedicine, econometrics, and political science. Researchers have utilized this framework to analyze various types of `expert' densities, including right-censored data \citep{Rosen1999MoE}, ranked preference data \citep{Gormley2008MoE, Gormley2011MoE}, and time-series \citep{Carvalho2007MoE}. We illustrate a graphical model representation of a full MoE and its restricted versions in Figure \ref{fig:MoE} following \citet{Gormley2011MoE}. 

\figurehere{1}

In the figure, $\boldsymbol{y}_{i}$ and $\boldsymbol{x}_{i}$ are the outcome variable and independent covariate for the $i^{th}$ individual, respectively, and $z_{i}$'s is the mixing component parameter (i.e., a latent categorical variable) of the $i^{th}$ individual. Additionally, $\boldsymbol{\beta_{g}}$ and $\boldsymbol{\beta_{e}}$ are `gating' (indirect) coefficients and `expert' (direct) coefficients, respectively. The difference between the full MoE and three possible reduced versions lies in the presence or absence of edges between the covariate $\boldsymbol{x}_{i}$ and the mixing component $z_{i}$ or the outcome variable $\boldsymbol{y}_{i}$. We interpret these models and link them to the corresponding counterpart in the structural equation modeling (SEM) literature if the equivalent model exists. 

\begin{enumerate}[label = (\alph*), topsep = 0pt, itemsep = -1ex, partopsep = 1ex, parsep = 1ex]
\item{In the finite mixture model (FMM, \citet{Muthen1999GMM}), the outcome variable $\boldsymbol{y}_{i}$ depends only on the mixing component parameter $z_{i}$. We express a FMM with $K$ latent classes as
\begin{equation}\nonumber
p(\boldsymbol{y}_{i})=\sum_{k=1}^{K}g(z_{i}=k)p(\boldsymbol{y}_{i}|\boldsymbol{\Theta}^{(k)}),
\end{equation}
where $g(z_{i}=k)$ is the proportion of the samples in cluster $k$ with two constraints $0\le g(z_{i}=k)\le 1$ and $\sum_{k=1}^{K}g(z_{i}=k)=1$, and $\boldsymbol{\Theta}^{(k)}$ is a set of class-specific parameters. The FMM has received lots of attention over the past twenty years in the SEM literature, with a considerable amount of empirical and theoretical work examining its benefits and limitations (for example, \citet{Bauer2003GMM, Muthen2004GMM, Grimm2009FMM, Nylund2007number, Grimm2010FMM}). Researchers usually employ the FMM to investigate sample heterogeneity and group individuals into subgroups that are more homogeneous \citep{Muthen2000GMM}.}
\item {In the gating-network mixture-of-experts model, the outcome variable $\boldsymbol{y}_{i}$ depends on the mixing component variable $z_{i}$ and the distribution of $z_{i}$ depends on covariates $\boldsymbol{x}_{i}$. Then we write a gating-expert MoE model with $K$ latent classes as
\begin{equation}\nonumber
p(\boldsymbol{y}_{i})=\sum_{k=1}^{K}g(z_{i}=k|\boldsymbol{x}_{gi})p(\boldsymbol{y}_{i}|\boldsymbol{\Theta}^{(k)}),
\end{equation}
where $g(z_{i}=k|\boldsymbol{x}_{gi})$, a multinomial function of covariates, is the proportion of the samples in cluster $k$, and has two constraints $0\le g(z_{i}=k|\boldsymbol{x}_{gi})\le 1$ and $\sum_{k=1}^{K}g(z_{i}=k|\boldsymbol{x}_{gi})=1$. The gating-expert MoE model is also popular among researchers who apply SEM. Previous studies have shown that including covariates in the gating functions can be realized in a confirmatory way through one-step models \citep{Clogg1981one, Goodman1974one, Haberman1979one, Hagenaars1993one, Vermunt1997one, Bandeen1997one, Dayton1988one, Kamakura1994one, Yamaguchi2000one} or in an exploratory fashion through two-step models \citep{Bakk2017two, Liu2019BLSGMM} or three-step models \citep{Clogg1995three, Bolck2004three, Vermunt2010three, Asparouhov2014three}. }
\item{In the expert-network mixture-of-experts model, the outcome variable $\boldsymbol{y}_{i}$ depends on both latent component membership $z_{i}$ and covariates $\boldsymbol{x}_{i}$; yet the distribution of the mixing component variable is independent of the covariates. So an expert-network MoE model with $K$ clusters is given
\begin{equation}\nonumber
p(\boldsymbol{y}_{i})=\sum_{k=1}^{K}g(z_{i}=k)p(\boldsymbol{y}_{i}|\boldsymbol{x}_{ei}, \boldsymbol{\Theta}^{(k)}), 
\end{equation}
which has the same constraints as the finite mixture model. To our knowledge, although the expert-network MoE model has not yet received much attention among researchers employing SEM, multiple existing studies in the SEM literature, for example, \citet{Asparouhov2014three, Masyn2017expert, Kim2016expert} have shown that the consequences of ignoring the direct effects that covariates have on sample heterogeneity can be severe by simulation studies}. Conceptually, it can be viewed as a mixture of multiple-indicator and multiple-cause (MIMIC) models in the SEM framework \citep{Joreskog1975MIMIC, McArdle1987MIMIC}. Each MIMIC model has two components: (1) a measurement model in which exogenous variables indicate latent variables; and (2) a structural model where covariates are multiple-causal predictors to the latent variables.
\item{In the full mixture-of-experts model, the outcome variable $\boldsymbol{y}_{i}$ depends on both the mixing component variable $z_{i}$ and covariates $\boldsymbol{x}_{i}$. Additionally, the distribution of the latent categorical variable $z_{i}$ also depends on the covariates $\boldsymbol{x}_{i}$. We then give a full MoE model with $K$ clusters as
\begin{equation}\nonumber
p(\boldsymbol{y}_{i})=\sum_{k=1}^{K}g(z_{i}=k|\boldsymbol{x}_{gi})p(\boldsymbol{y}_{i}|\boldsymbol{x}_{ei}, \boldsymbol{\Theta}^{(k)}),
\end{equation}
which has the same constraints as the gating-network MoE model. The full MoE model is not a brand-new concept in the SEM framework. When introducing the FMM into the SEM framework, \citet{Muthen1999GMM} viewed it as a possible generalization of the finite mixture model. It can also be viewed as an extension of the expert-network MoE by allowing the latent component membership of MIMIC components to be multinomial functions of covariates. Additionally, \citet{Asparouhov2014three} have examined the impact of direct and indirect effects of covariates in the context of growth mixture models where the direct effect is on the growth factors.} Note that the covariates of the mixing components variable $z_{i}$ and the outcome variable $\boldsymbol{y}_{i}$ can be the same or different.
\end{enumerate}

In short, the full MoE and its three reduced models all have their corresponding counterparts in the SEM framework: the full MoE model is a full mixture model with consideration of both direct and indirect effects that covariates have on sample heterogeneity. The expert-network MoE, the gating-network MoE, and the FMM are all restricted models obtained by fixing logistic coefficients (i.e., effects of covariates with indirect effects), path coefficients (i.e., effects of covariates with direct effects), and both to be zero. In the following text, we mainly use SEM terminology, as it is more familiar to social science researchers, and refer to the full MoE model, the expert-network MoE model, and the gating-network MoE model as the full mixture model, growth predictor mixture (GP-mixture) model, and cluster predictor mixture (CP-mixture) model, respectively. However, the model labels may be considered interchangeable.

A within-class model can take multiple forms in the SEM framework. For example, it can be a factor model with covariates, where the latent variables are indicated by the outcome variable $\boldsymbol{y}_{i}$ and caused by covariates $\boldsymbol{x}_{i}$ directly. It can also be a latent growth curve model with time-invariant covariates (LGC-TICs), where the latent variables are growth factors indicated by the repeated measures of $\boldsymbol{y}_{i}$ and directly caused by covariates $\boldsymbol{x}_{i}$. More importantly, researchers can specify the parameters that need to be fixed or freely estimated in each class. For example, in a two-class full mixture model with LGC-TICs, the underlying functional form of trajectories could be specified as quadratic in the first class but linear such that the mean and variance of the quadratic term as well as quadratic-intercept, quadratic-linear covariances are fixed to zero in the second class. 

This article focuses on a full mixture model with a nonlinear LGC-TICs as the within-class model. Specifically, we consider a bilinear growth model \citep[Chapter~11]{Grimm2016growth} (see Figure \ref{fig:E2_2cases}), also referred to as a linear-linear piecewise model \citep{Harring2006nonlinear, Kohli2011PLGC, Kohli2013PLGC1, Kohli2013PLGC2, Kohli2015PLGC1}, with an unknown knot and time-invariant covariates (TICs) in each latent class. We decide to employ the bilinear spline growth model with TICs (BLSGM-TICs) as the within-class model for two reasons. On the one hand, this piecewise functional form allows for investigating the change of growth rate during different developmental stages and when the change of growth rate occurs, which are often of interest in developmental studies. On the other hand, \citet{Liu2019BLSGM} has shown that the bilinear spline functional form is a better fit for repeated mathematics scores in our motivating data set than models with parametric functions such as quadratic and Jenss-Bayley. Following multiple existing studies, for example, \citet{Sterba2014individually, Preacher2015repara, Liu2019BLSGMM}, we build the model in the framework of individual-measurement occasions by using `definition variables' (observed variables that adjust model parameters to individual-specific values) \citep{Mehta2000people, Mehta2005people} to avoid possible inadmissible solutions \citep{Blozis2008coding, Coulombe2015ignoring}. 

\subsection{Challenges of Mixture-of-Experts Models Implementation}
Unless a study is conducted to answer a specific question, we usually have two challenges when specifying a full mixture model, deciding the number of clusters and selecting which covariates, if any, need to be included, and if so, in which cluster(s) or multinomial function(s). Earlier studies have proposed approaches to decide the number of latent classes for different types of within-class models. For example, \citet{Jacobs1997number} addressed this issue for a within-class model that is a generalized linear model, \citet{Rosen2000number} handled this issue in the context of marginal models, and \citet{Rosen1999MoE} developed an approach to decide the number of clusters for a mixture of proportional hazards models. Additionally, \citet{Zeevi1998number, Wood2002number} and \citet{Carvalho2005number} advocated for using a penalized criterion, such as Akaike Information Criterion (AIC), Bayesian Information Criterion (BIC), or Minimum Description Length (MDL) criterion, to choose the number of clusters in the model. 

In the SEM framework, \citet{Diallo2017number} and \citet{Nylund2016number} have conducted extensive simulation studies to investigate the impact of covariates with direct and/or indirect effects on the latent class enumeration process in the context of the multilevel growth mixture modeling framework and cross-sectional latent class models with binary indicators, respectively. Both studies have recommended conducting the enumeration process without the inclusion of covariates and employing the BIC to determine the optimal number, especially under conditions of large class separation \citep{Diallo2017number}. In the current study, we do not intend to develop a novel metric for choosing the number of clusters in the context of full growth mixture models. Instead, we follow the convention in the SEM literature to determine the number in an exploratory fashion. We fit a pool of candidate finite mixture models (i.e., not to include any covariates) with different numbers of latent classes and pick the `best' model along with the desired number of clusters via statistical criteria such as the BIC \citep{Nylund2007number}.

Another challenge of specifying a full mixture model is to decide which covariates should be added to the model, especially in the SEM framework. The number of potential covariates could be large, and highly-correlated covariate subsets may exist in the psychological and educational domains, where the SEM framework is widely utilized. In the SEM literature, examining the relationship between latent classes and covariates can be realized by the one-step approach \citep{Bandeen1997one} or stepwise methods. The measurement parameters and the coefficients from the predictors to the latent class variable are estimated simultaneously in the one-step approach. Multiple existing studies, for example, \citet{Asparouhov2014three, Bakk2017two, Kim2016expert, Hsiao2020mediation}, have shown that the one-step approach usually performs better than those stepwise methods in terms of performance metrics such as bias, RMSE and coverage probability, although \citet{Vermunt2010three} pointed out several disadvantages of the one-step approach in the context of covariates of the latent class variable. The major critiques lie in (1) computational burden when the dimension of the potential covariates is large, especially in a more exploratory study, and (2) whether or not to include covariates when deciding the number of latent classes. Multiple studies, for example, \citet{Clogg1995three, Vermunt2010three, Bolck2004three, Bakk2017two}, have proposed stepwise methods to avoid such drawbacks. Although these methods are different in the procedure, their primary idea is the same: to separate the estimation of measurement parameters and coefficients in the multinomial functions.

Researchers have recently recommended employing the adjusted one-step approach \citep{Kim2016expert, Hsiao2020mediation} to address the second critique. In the adjusted one-step approach, a stable solution for enumeration is determined with no covariates (i.e., the process we stated earlier for this study). Then the class-specific parameters and logistic coefficients are estimated simultaneously with the determined number of clusters. In this study, we propose a possible approach that can help identify the most important covariates among a set of candidates efficiently by leveraging machine learning techniques (i.e., address the first critique on the one-step model). Specifically, we propose to employ structural equation model forests (SEM Forests, \citet{Brandmaier2016semForest}), an extension of random forests (RFs, \citet{Breiman2001forest}) in the SEM framework, to select covariates. Note that we aim to introduce how to use its output named `variable importance' with a basic understanding of its algorithm instead of examining this method comprehensively.

Both RFs and SEM Forests have their `simple-tree' versions: classification and regression trees (CARTs, \citet{Breiman1984CARTs}) and structural equation model trees (SEM Trees, \citet{Brandmaier2013semTree}). CARTs can be utilized to analyze univariate continuous outcomes (regression trees) or categorical outcomes (classification trees). The algorithm that lies behind CARTs is intuitive: it regresses the outcome variable on a set of candidate covariates and starts with an empty tree. At each step, the algorithm needs to select a covariate to split and the value of the threshold (of a continuous covariate or a categorical covariate with more than two levels) to optimize a pre-specified metric (for example, to minimize the sum-of-squares errors in a regression problem or maximize accuracy in a classification problem). The algorithm usually conducts this optimization by exhaustive search and does not stop this partition process until the sample homogeneity of each (sub-)split cannot be improved further \citep[Chapter~ 14]{Bishop2006pattern}. By extending the CARTs from a univariate outcome setting to the scenario with a multivariate outcome variable, SEM Trees expedite exploratory analyses in the SEM framework \citep{Brandmaier2013semTree}. One available objective metric to be optimized in the SEM Trees is the likelihood function. The algorithm selects one covariate to partition, along with a selected threshold value, to maximize the likelihood function of this split.

Both CARTs and SEM Trees suffer an overfitting issue that is an inherent limitation of the algorithm. The algorithm is designed to optimize a metric, say accuracy or likelihood function, greedily for one sample so that the fit model cannot be generalized to other samples from the same population. Other than remedies such as pruning \citep{Breiman1984CARTs, Brandmaier2013semTree}, an ensemble or `bagging' technique proposed as RFs \citep{Breiman2001forest} and SEM Forests \citep{Brandmaier2016semForest} also addresses the overfitting issue for CARTs and SEM Trees, respectively. As the word `forests' suggests, both a RF and a SEM Forest are a collection of trees constructed by resampling with randomly selected covariates of the original dataset. A forest is more stable than a single tree as it is an average of all constructed trees \citep{Breiman2001forest, Brandmaier2016semForest}. More importantly, both forest algorithms calculate variable importance to quantify the (relative) impact a covariate has on the metric. The variable importance obtained from the SEM Forest is a score to assess how important a covariate is in predicting the multivariate means and variance-covariance structure. Only covariates with high importance or effects on sample heterogeneity require further examination. 

Although a covariate could have direct and indirect effects on sample heterogeneity simultaneously and it is realizable in the SEM framework by adding paths from the covariate to the latent categorical variable and to the class-specific growth factors, we allow the covariates in the within-class model to differ from those in the multinomial functions based on conceptual and technical considerations that we will explain in detail in the Discussion section. In this study, with the set of selected covariates, we place the ones that are assumed to directly affect sample heterogeneity in the latent classes to explain the within-class variability while the others in the multinomial functions to predict the membership of each individual. 

In the remainder of this article, we describe how to specify and estimate a full mixture model with the BLSGM-TICs as the within-class model. In the subsequent section, we depict the simulation design for model evaluation. We evaluate model performance through its estimates and clustering effects. We then demonstrate how to employ SEM Forests to select baseline covariates efficiently. In the Application section, we analyze the motivating data, longitudinal mathematics item response theory (IRT) scores, demonstrating how to implement the SEM Forests to select covariates and construct the proposed mixture models. Additionally, we demonstrate how to extend the proposed model in the SEM framework by allowing the covariates in latent classes to be time-varying. We finally frame discussions concerning the methodological and practical considerations as well as future directions.

\section{Method}\label{method}
\subsection{Model Specification of the Full Mixture Model}\label{method:spec}
In this section, we specify the proposed full mixture model in the SEM framework. Specifically, we assume that the within-class model takes the functional form of a bilinear spline growth curve with an unknown knot for the underlying change patterns. In each latent class, we also add covariates to explain the within-class variability of trajectories (growth factors). \citet{Harring2006nonlinear} pointed out there are five parameters in the linear-linear piecewise functional form: an intercept and slope of each linear piece and a knot, but the degree-of-freedom of the bilinear spline reduces to four as two linear pieces join at the knot. In the current study, we consider the initial intercept, two slopes, and the knot as the four free parameters following multiple existing studies, for example, \citet{Kohli2011PLGC, Kohli2013PLGC1, Kohli2013PLGC2}. Although \citet{Preacher2015repara, Liu2019BLSGM, Liu2020PBLSGM} have shown that the knot can be an additional growth factor with considering its variability, we construct a parsimonious model assuming that the class-specific knot is at the same time point for all individuals in each cluster as the knot variability is not the aim of the current study. For the $i^{th}$ individual, we express the model in the SEM framework as
\begin{align}
&p(\boldsymbol{y}_{i}|z_{i}=k,\boldsymbol{x}_{gi}, \boldsymbol{x}_{ei})=\sum_{k=1}^{K}g(z_{i}=k|\boldsymbol{x}_{gi})\times p(\boldsymbol{y}_{i}|z_{i}=k,\boldsymbol{x}_{ei}),\label{eq:MoE_full}\\
&g(z_{i}=k|\boldsymbol{x}_{gi})=\begin{cases}
\frac{1}{1+\sum_{k=2}^{K}\exp(\beta_{g0}^{(k)}+\boldsymbol{\beta}_{g}^{(k)T}\boldsymbol{x}_{gi})}& \text{Reference Group ($k=1$)}\\
\frac{\exp(\beta_{g0}^{(k)}+\boldsymbol{\beta}_{g}^{(k)T}\boldsymbol{x}_{gi})} {1+\sum_{k=2}^{K}\exp(\beta_{g0}^{(k)}+\boldsymbol{\beta}_{g}^{(k)T}\boldsymbol{x}_{gi})} & \text{Other Groups ($k=2,\dots, K$)}
\end{cases},\label{eq:Gating_full}\\
&\boldsymbol{y}_{i}|(z_{i}=k,\boldsymbol{\eta}_{i})=\boldsymbol{\Lambda}_{i}(\gamma^{(k)})\boldsymbol{\eta}_{i}|(z_{i}=k,\boldsymbol{x}_{ei})+\boldsymbol{\epsilon}_{i}|(z_{i}=k), \label{eq:Experts1}\\
&\boldsymbol{\eta}_{i}|(z_{i}=k,\boldsymbol{x}_{ei})=\boldsymbol{\beta}_{e0}^{(k)}+\boldsymbol{\beta}_{e}^{(k)}\boldsymbol{x}_{ei}|(z_{i}=k)+\boldsymbol{\zeta}_{i}|(z_{i}=k). \label{eq:Experts2_full}
\end{align}

Equation (\ref{eq:MoE_full}) defines a probabilistic model that combines mixing proportions, $g(z_{i}=k|\boldsymbol{x}_{gi})$, and within-class models, $p(\boldsymbol{y}_{i}|z_{i}=k,\boldsymbol{x}_{ei})$, where $\boldsymbol{x}_{gi}$, $\boldsymbol{x}_{ei}$, $\boldsymbol{y}_{i}$ and $z_{i}$ are the covariates with indirect effects, covariates with direct effects, $J\times1$ vector of repeated outcomes (in which $J$ is the number of measurements) and the membership of individual $i$, respectively. Note that there are two constraints on Equation (\ref{eq:MoE_full}): $0\le g(z_{i}=k|\boldsymbol{x}_{gi})\le 1$ and $\sum_{k=1}^{K}g(z_{i}=k|\boldsymbol{x}_{gi})=1$. Equation (\ref{eq:Gating_full}) defines mixing components as multinomial functions of covariates $\boldsymbol{x}_{gi}$, where $\beta_{g0}^{(k)}$ and $\boldsymbol{\beta}_{g}^{(k)}$ are the intercept and the logistic coefficients that indicate indirect effects, respectively. The multinomial functions decide which cluster an individual belongs to with uncertainty, depending on the values of $\boldsymbol{x}_{gi}$. 

Equations (\ref{eq:Experts1}) and (\ref{eq:Experts2_full}) together define a within-class model. Equation (\ref{eq:Experts1}) writes the outcome $\boldsymbol{y}_{i}$ as a linear combination of growth factors $\boldsymbol{\eta}_{i}$. When the within-class model takes the bilinear spline functional form with an unknown knot, $\boldsymbol{\eta}_{i}$ is a $3\times1$ vector of growth factors ($\boldsymbol{\eta}_{i}=\eta_{0i},\eta_{1i},\eta_{2i}$, for an intercept and two slopes); accordingly, $\boldsymbol{\Lambda}_{i}(\gamma^{(k)})$ is a $J\times3$ matrix of factor loadings. The subscript $i$ in $\boldsymbol{\Lambda}_{i}(\gamma^{(k)})$ accounts for individual measurement occasions. Note that $\boldsymbol{\Lambda}_{i}(\gamma^{(k)})$ is a function of the class-specific knot $\gamma^{(k)}$. In our study, the knot can occur at a particular measurement occasion or between two measurements. The repeated outcomes $\boldsymbol{y}_{i}$ have different pre- and post-knot expressions
\begin{equation}\nonumber
y_{ij}=\begin{cases}
\eta_{0i}+\eta_{1i}t_{ij}+\epsilon_{ij} & t_{ij}\le\gamma^{(k)}\\
\eta_{0i}+\eta_{1i}\gamma^{(k)}+\eta_{2i}(t_{ij}-\gamma^{(k)})+\epsilon_{ij} & t_{ij}>\gamma^{(k)}\\
\end{cases}, 
\end{equation}
where $y_{ij}$ and $t_{ij}$ are the measurement and its occasion of the $i^{th}$ individual at time $j$.

Equation (\ref{eq:Experts2_full}) further regresses the growth factors $\boldsymbol{\eta}_{i}$ on covariates with direct effects, where $\boldsymbol{\beta}_{e0}^{(k)}$ and $\boldsymbol{\beta}_{e}^{(k)}$ are a $3\times1$ vector of growth factor intercepts and a $3\times c$ matrix of path coefficients (where $c$ is the number of covariates) in cluster $k$, respectively. Additionally, $\boldsymbol{x}_{ei}|(z_{i}=k)$ is a $c\times1$ vector of covariates, and $\boldsymbol{\zeta}_{i}|(z_{i}=k)$ is a $3\times1$ vector of deviations of the $i^{th}$ individual from the class-specific growth factors conditional means. 

\citet[Chapter~11]{Grimm2016growth}, \citet{Harring2006nonlinear, Preacher2015repara, Liu2019BLSGM} presented multiple ways to unify pre- and post-knot expressions by reparameterizing growth factors. In this article, we follow \citet{Liu2019BLSGM} as the transformation function and matrix for the growth factors and corresponding path coefficients between the original and reparameterized frames are available for this reparameterizing strategy. Note that the expressions of the repeated outcome $\boldsymbol{y}_{i}$ in two parameter-spaces are mathematically equivalent, although only the parameters in the original frame are directly related to the underlying change patterns. We provide the reparameterizing process and details of transformation between class-specific growth factors and path coefficients of the two parameter-spaces in Appendices \ref{Supp:1A} and \ref{Supp:1B}.

\subsection{Model Estimation}\label{method:est}
To simplify the model, we make two assumptions. (1) Within-class growth factors follow a multivariate Gaussian distribution conditional on covariates, that is, the vector of deviations $\boldsymbol{\zeta}_{i}|(z_{i}=k)\sim \text{MVN}(\boldsymbol{0}, \boldsymbol{\Psi}^{(k)}_{\boldsymbol{\eta}})$, where $\boldsymbol{\Psi}^{(k)}_{\boldsymbol{\eta}}$ is the unexplained variance-covariance matrix of class-specific growth factors. (2) Individual residuals are independent and identically normally distributed over time in each latent class, that is, $\boldsymbol{\epsilon}_{i}|(z_{i}=k)\sim N(\boldsymbol{0}, \theta_{\epsilon}^{(k)}\boldsymbol{I})$, where $\boldsymbol{I}$ is a $J\times J$ identity matrix. Accordingly, the within-class implied mean vector ($\boldsymbol{\mu}_{i}^{(k)}$) and variance-covariance matrix ($\boldsymbol{\Sigma}_{i}^{(k)}$) of repeated outcomes $\boldsymbol{y}_{i}$ for the $i^{th}$ individual in the $k^{th}$ cluster are given as
\begin{align}
&\boldsymbol{\mu}_{i}^{(k)}=\boldsymbol{\Lambda}_{i}(\gamma^{(k)})(\boldsymbol{\beta}_{e0}^{(k)}+\boldsymbol{\beta}_{e}^{(k)}\boldsymbol{\mu}_{\boldsymbol{x}e}^{(k)}),\label{eq:Ymean_full}\\
&\boldsymbol{\Sigma}_{i}^{(k)}=\boldsymbol{\Lambda}_{i}(\gamma^{(k)})\boldsymbol{\Psi}^{(k)}_{\boldsymbol{\eta}}\boldsymbol{\Lambda}_{i}(\gamma^{(k)})^{T}+\boldsymbol{\Lambda}_{i}(\gamma^{(k)})\boldsymbol{\beta}_{e}^{(k)}\boldsymbol{\Phi}^{(k)} \boldsymbol{\beta}_{e}^{(k)T}\boldsymbol{\Lambda}_{i}(\gamma^{(k)})^{T}+\theta_{\epsilon}^{(k)}\boldsymbol{I}\label{eq:Yvar_full},
\end{align}
respectively, where $\boldsymbol{\mu}_{\boldsymbol{x}e}^{(k)}$ and $\boldsymbol{\Phi}^{(k)}$ are the mean vector ($c\times 1$) and the variance-covariance matrix ($c\times c$) of the covariates in the $k^{th}$ latent class, respectively.

The parameters that need to be estimated in the proposed model include
\begin{equation}\label{eq:theta4}
\begin{aligned}
\boldsymbol{\Theta}_{\text{full}}&=\{\boldsymbol{\beta}_{e0}^{(k)}, \boldsymbol{\Psi}^{(k)}_{\boldsymbol{\eta}}, \gamma^{(k)}, \boldsymbol{\beta}_{e}^{(k)}, \boldsymbol{\mu}_{\boldsymbol{x}e}^{(k)}, \boldsymbol{\Phi}^{(k)}, \theta_{\epsilon}^{(k)}, \beta_{g0}^{(k)}, \boldsymbol{\beta}_{g}^{(k)}\}\\
&=\{\beta_{\eta_{0}}^{(k)}, \beta_{\eta_{1}}^{(k)}, \beta_{\eta_{2}}^{(k)}, \psi_{00}^{(k)}, \psi_{01}^{(k)}, \psi_{02}^{(k)}, \psi_{11}^{(k)}, \psi_{12}^{(k)}, \psi_{22}^{(k)}, \gamma^{(k)}, \boldsymbol{\beta}_{e}^{(k)}, \boldsymbol{\mu}_{\boldsymbol{x}e}^{(k)}, \boldsymbol{\Phi}^{(k)}, \theta_{\epsilon}^{(k)}, \beta_{g0}^{(k)}, \boldsymbol{\beta}_{g}^{(k)}\},\\
&k=2,\dots,K \text{ for } \beta_{g0}^{(k)}, \boldsymbol{\beta}_{g}^{(k)},\\
&k=1,\dots,K \text{ for other parameters}.
\end{aligned}
\end{equation}

The log-likelihood function of the model specified in Equations (\ref{eq:MoE_full})-(\ref{eq:Experts2_full}) is
\begin{equation}\label{eq:lik}
\begin{aligned}
\ell\ell(\boldsymbol{\Theta}_{\text{full}})&=\sum_{i=1}^{n}\log\bigg(\sum_{k=1}^{K}g(z_{i}=k|\boldsymbol{x}_{gi})p(\boldsymbol{y}_{i}|z_{i}=k,\boldsymbol{x}_{ei})\bigg)\\
&=\sum_{i=1}^{n}\log\bigg(\sum_{k=1}^{K}g(z_{i}=k|\boldsymbol{x}_{gi})p(\boldsymbol{y}_{i}|\boldsymbol{\mu}_{i}^{(k)},\boldsymbol{\Sigma}_{i}^{(k)},\boldsymbol{x}_{ei})\bigg).
\end{aligned}
\end{equation}

Multiple techniques are available to estimate parameters in mixture models. In the machine learning literature, one recommended approach is the expectation-maximization (EM) algorithm as the mixing component $z_{i}$ in Equation (\ref{eq:lik}) is unknown and the EM algorithm gets around this problem by viewing it as known with an initial guess and updating it at each iteration until achieving convergent status. We first define the cluster responsibilities (i.e., posterior probabilities) of an iteration $t$ as
\begin{equation}\label{eq:response_full}
\hat{r}_{ik}^{(t)}=\frac{g(z_{i}=k|\boldsymbol{x}_{gi})^{(t-1)}p(\boldsymbol{y}_{i}|\hat{\boldsymbol{\mu}}_{i}^{(k)(t-1)}, \hat{\boldsymbol{\Sigma}}_{i}^{(k)(t-1)}, \boldsymbol{x}_{ei})}{\sum_{k=1}^{K}g(z_{i}=k|\boldsymbol{x}_{gi})^{(t-1)}p(\boldsymbol{y}_{i}|\hat{\boldsymbol{\mu}}_{i}^{(k) (t-1)}, \hat{\boldsymbol{\Sigma}}_{i}^{(k) (t-1)}, \boldsymbol{x}_{ei})}.
\end{equation}
In each iteration, the EM algorithm consists of two steps: E-step, which estimates cluster responsibilities given current parameter estimates, and M-step, which maximizes the likelihood over parameters given current responsibilities. More technical details about the EM algorithm can be found in \citet[Chapter~11]{Murphy2012MoE}. 

The EM algorithm is available in multiple SEM software such as \textit{Mplus} and the \textit{R} package \textit{OpenMx}. By specifying `algorithm=EM', we can request a pure EM algorithm in \textit{Mplus}. In the \textit{R} package \textit{OpenMx}, a model is optimized by the EM algorithm when we specify E-step and M-step in the computation plan \citep{OpenMx2016package, Pritikin2015OpenMx, Hunter2018OpenMx, User2020OpenMx}. Additionally, the default optimizer \textit{CSOLNP} of \textit{OpenMx}, which utilizes the Broyden-Fletcher-Goldfarb-Shanno (BFGS) algorithm (an iterative method that belongs to quasi-Newton methods for addressing unconstrained nonlinear optimization problems) has been shown efficiently for mixture models in an existing study \citep{Liu2019BLSGMM}. In this study, we use the \textit{CSOLNP} optimizer to estimate the parameters listed in Equation (\ref{eq:theta4}) and provide the \textit{OpenMx} syntax along with a demonstration in the online appendix (\url{https://github.com/Veronica0206/Extension_projects}). The \textit{Mplus} 8 code for the proposed model is also provided in the online appendix. (The code will be uploaded upon acceptance.)

We can specify three possible restricted models through removing the paths between the membership $z_{i}$ and the covariates $\boldsymbol{x}_{i}$, that between the covariates $\boldsymbol{x}_{i}$ and the outcome variable $\boldsymbol{y}_{i}$ and both from the full model defined in Equations (\ref{eq:MoE_full})-(\ref{eq:Experts2_full}). The mean-vector and variance-covariance matrix of repeated outcomes in Equations (\ref{eq:Ymean_full}) and (\ref{eq:Yvar_full}), the parameters in Equation (\ref{eq:theta4}) and the defined cluster responsibilities at an iteration $t$ in Equation (\ref{eq:response_full}) need to be updated accordingly. We provide these equations in Appendix \ref{Supp:1C}.

\section{Model Evaluation}\label{Evaluation}
The proposed mixture model with the BLSGM-TICs as the within-class model and its three possible reduced versions are evaluated using a Monte Carlo simulation study with three goals. In the simulation study, we consider two covariates with indirect effects and two covariates with direct effects. The first goal is to examine the performance measures of the proposed models and clustering effects when we specify them correctly, no matter in the full or any restricted form. The performance metrics include the relative bias, empirical standard error (SE), relative root-mean-square error (RMSE), and empirical coverage for a nominal 95\% confidence interval (CI) of each parameter. Definitions and estimates of these four performance measures are listed in Table \ref{tbl:metric}. 

\tablehere{1}

To evaluate the clustering effects, we first need to calculate the posterior probability for each individual belonging to each cluster as Equations (\ref{eq:ent_full}), (\ref{eq:ent_expert}), (\ref{eq:ent_gating}) and (\ref{eq:ent_FMM}) for full mixture models, GP-mixture models, CP-mixture models and FMMs:
\begin{align}
&p(z_{i}=k)=\frac{g(z_{i}=k|\boldsymbol{x}_{gi})p(\boldsymbol{y}_{i}|z_{i}=k, \boldsymbol{x}_{ei})}{\sum_{k=1}^{K}g(z_{i}=k|\boldsymbol{x}_{gi})p(\boldsymbol{y}_{i}|z_{i}=k, \boldsymbol{x}_{ei})}\label{eq:ent_full},\\
&p(z_{i}=k)=\frac{g(z_{i}=k)p(\boldsymbol{y}_{i}|z_{i}=k, \boldsymbol{x}_{ei})}{\sum_{k=1}^{K}g(z_{i}=k)p(\boldsymbol{y}_{i}|z_{i}=k, \boldsymbol{x}_{ei})}\label{eq:ent_expert},\\
&p(z_{i}=k)=\frac{g(z_{i}=k|\boldsymbol{x}_{gi})p(\boldsymbol{y}_{i}|z_{i}=k)}{\sum_{k=1}^{K}g(z_{i}=k|\boldsymbol{x}_{gi})p(\boldsymbol{y}_{i}|z_{i}=k)}\label{eq:ent_gating},\\
&p(z_{i}=k)=\frac{g(z_{i}=k)p(\boldsymbol{y}_{i}|z_{i}=k)}{\sum_{k=1}^{K}g(z_{i}=k)p(\boldsymbol{y}_{i}|z_{i}=k)}\label{eq:ent_FMM}.
\end{align}

With each individual's posterior probabilities vector, we assign the individual to the cluster with the highest posterior probability. The tie among competing components with equally maximum probabilities is randomly broken as described in \citet{McLachlan2000FMM}. The clustering effects include accuracy and entropy. As we have true membership in a simulation study, accuracy, which is defined as the fraction of all correctly classified instances, is available to assess how the algorithm separates the samples into `correct' groups \citep[Chapter~ 1]{Bishop2006pattern}. Entropy is a metric based on the average posterior probabilities \citep{Stegmann2018Entropy}, which is given as
\begin{equation}\label{eq:ent}
\text{Entropy}=1+\frac{1}{n\log(K)}\bigg(\sum_{n=1}^{n}\sum_{k=1}^{K}p(z_{i}=k)\log p(z_{i}=k)\bigg).
\end{equation}
It measures the uncertainty of the clustering results and ranges from $0$ to $1$, with $0$ and $1$ indicating no class separation and complete class separation, respectively. Earlier studies have shown that entropy is a good indicator of accuracy if we can correctly specify the covariates. For example, \citet{Lubke2007entropy} have shown that entropy values around $0.8$ or above suggest at least $90\%$ accuracy, although the cutoff of entropy could be sensitive to the within-class model.

The second goal is to see whether placing all four covariates in the multinomial functions would result in inadmissible solutions or misleading information. It is common misspecification since researchers usually assume that the covariates only have indirect effects on sample heterogeneity. With this misspecified model, we also aim to examine the performance of entropy when the constructed model includes model misspecification. The third goal is to compare the four correctly-specified models\footnote{In this context, the correctly-specified models are broadly defined (including the full mixture model and its restricted models), suggesting that we did not misspecify (though could under-specify) the covariates.}, among themselves and with the misspecified model. We first examine Dumenci's Latent Kappa \citep{Dumenci2011kappa, Dumenci2019knee} between the trajectory clusters obtained from the FMM and those from each mixture models with covariates. We also investigate whether we can apply common statistical criteria, such as the BIC, to select the `best' model among the proposed mixture model and its three restricted models. 

In the simulation design, we decided the number of repetitions $S=1,000$ by an empirical approach following \citet{Morris2019simulation}. We conducted a pilot study and observed that standard errors of all parameters except the intercept (unexplained) variances were less than $0.15$. Bias is the most important performance metric. To keep the Monte Carlo standard error of bias\footnote{$\text{Monte Carlo SE(Bias)}=\sqrt{Var(\hat{\theta})/S}$ \citep{Morris2019simulation}.} for the majority of parameters (i.e., all parameters except the intercept variances) lower than $0.005$, we needed at least $900$ replications. We then decided to proceed with $S=1,000$ to be more conservative. 

\subsection{Design of Simulation Study}\label{Evaluation:design}
Table \ref{tbl:simu} lists all conditions that we considered in the simulation design. We fixed the conditions, including the sample size, the number of clusters, the number of repeated measurements, the variance-covariance matrix of the growth factors, the distribution of the covariates with direct effects, and the time-window of individual measurement occasions, which are not of primary interest in this study. For example, we selected ten scaled and equally spaced waves since \citet{Liu2019BLSGM, Liu2020PBLSGM} have shown that bilinear growth models performed decently in terms of performance measures under this condition and that fewer numbers (for example, six) of repeated outcomes only affected model performance slightly. Similar to earlier studies, we allowed for a moderate time-window, $(-0.25, +0.25)$, of individual measurement occasions around each wave \citep{Coulombe2015ignoring}. Since the variance-covariance structure of the growth factors usually changes with the measurement scales and time scales, we fixed it and kept the index of dispersion ($\sigma^{2}/\mu$) of each growth factor at one-tenth scale to follow \citet{Bauer2003GMM, Kohli2011PLGC, Kohli2015PLGC1}. The correlations between growth factors were set to be a moderate level ($\rho=0.3$). 

\tablehere{2}

The most important feature of a model-based clustering algorithm lies in how well it can detect sample heterogeneity and estimate parameters of interest. Intuitively, the model should perform better under conditions with larger separation between latent classes. The distance between latent classes is measured by the difference between the class-specific density functions. Therefore, there are two possible metrics to gauge the separation between two clusters: the Mahalanobis distance between class-specific growth factors and the difference in the knot locations \citep{Kohli2015PLGC1, Liu2019BLSGMM}. In this study, we kept the Mahalanobis distance as $0.86$ (that is a small distance as defined by \citet{Kohli2015PLGC1}) and set $1.0$, $1.5$ and $2.0$ as a small, medium and large difference in knot locations. Those manipulated conditions allowed us to examine how the separation between the two latent classes affects model performance. Note that we set $\boldsymbol{x}_{e}$'s distribution to be the same across latent classes in this simulation study to avoid its possible influence on the distance between clusters. Additionally, we considered two levels of allocation ratios: $1$:$1$ and $1$:$2$, roughly controlled by the intercept coefficients in the multinomial functions. We selected the class mixing proportion of $1$:$1$ as a balanced allocation; we chose the other level as we wanted to examine model performance in a more challenging scenario concerning mixing proportions. 

Another important feature that we wanted to investigate through the simulation study is how the two types of covariates affect the proposed model in terms of performance measures and clustering effects. We standardized all covariates so that the effect sizes of the same type of covariates are comparable. We fixed the coefficients of two covariates with indirect effects as $\log(1.5)$ (i.e., the odds ratio is $1.5$) and $\log(1.7)$, respectively. We adjusted the relative importance of the covariates with direct effects against those with indirect effects by varying path coefficients to account for the moderate ($13\%$) or substantial ($26\%$) variance of class-specific growth factors \citep[Chapter~9]{Cohen1988R}. The covariates can explain $13\%$,  $13\%$ and $26\%$ for within-class trajectory variability in one cluster and $13\%$, $26\%$ and $26\%$ variability in the other. Additionally, we considered two levels of residual variance to examine how measurement precision affects the model. We also considered three scenarios (Scenario 1, 2 and 3 in Table \ref{tbl:simu}) to see if the trajectory shape affects model performance. We varied the knot location and one growth factor while keeping the other two growth factors the same across the clusters in each scenario.

\subsection{Data Generation and Simulation Step}\label{evaluation:step}
We carried out two-step data generation for each condition in Table \ref{tbl:simu}. We first obtained the membership $z_{i}$ from covariates with indirect effects for each individual. We then generated the outcome variable $\boldsymbol{y}_{i}$ and covariates with direct effects for each cluster simultaneously. The general steps of data generation are: 
\begin{enumerate}
 \item Create membership $z_{i}$ for the $i^{th}$ individual:
 \begin{enumerate}
 \item Generate data matrix of covariates with indirect effects $\boldsymbol{x}_{g}$,
 \item Calculate the probability vector for each entry based on the covariates with indirect effects and a set of specified logistic coefficients with a multinomial link and assign each individual to the component with the highest probability,
 \end{enumerate} 
 \item Generate data of growth factors and covariates with direct effects $\boldsymbol{x}_{e}$ simultaneously for each component using the \textit{R} package \textit{MASS} \citep{Venables2002Statistics},
 \item Generate the time structure with ten scaled and equally-spaced waves $t_{j}$ and obtain individual measurement occasions by allowing the time-window as $t_{ij}\sim U(t_{j}-\Delta, t_{j}+\Delta)$ around each wave,
 \item Calculate factor loadings, which are functions of the class-specific knot and individual measurement occasions, for each individual,
 \item Calculate values of the repeated outcomes from the class-specific growth factors, corresponding factor loadings, class-specific knot and residual variances,
 \item Apply the proposed mixture models to the generated data, estimate the parameters, and construct corresponding $95\%$ Wald CIs, along with accuracy and entropy,
 \item Repeat Steps $1$-$6$ until achieving $1,000$ convergent solutions, calculate the relative bias, empirical SE, relative RMSE, and coverage probability of each parameter,
 \item Respecify a model with all covariates in the multinomial functions on the data sets from which we obtained convergent solutions in the above steps. 
\end{enumerate}

\section{Results}\label{results}
\subsection{Model Convergence}
We first examined the convergence\footnote{Convergence is defined as to achieve \textit{OpenMx} status code $0$, which suggests a successful optimization, until up to $10$ attempts with different collections of starting values \citep{OpenMx2016package}.} rate of the proposed full mixture model and its three reduced versions under each condition in this section. The convergence rate of the full, GP-, CP- and finite mixture model achieved at least $89\%$, $89\%$, $87\%$ and $87\%$, respectively across all conditions in the simulation study. Out of a total of $108$ conditions, $36$, $35$, $36$ and $35$ conditions reported $100\%$ of convergence rate for the full, GP-, CP- and finite mixture model, respectively. We noticed that all of these conditions with $100\%$ convergence rate were those with a large difference in knot locations (i.e., the knot location difference is $2$). Additionally, $54$, $54$, $51$ and $51$ conditions reported convergence rates of $95\%$ to $99\%$ for the full, GP-, CP- and finite mixture model, respectively. The worst scenario regarding the non-convergence rate is $153/1153$, indicating that we need to repeat the process described above $1,153$ times to have $1,000$ replications with a convergent solution. It occurred when we tried to fit a CP-mixture model under the conditions with balanced allocation, the small difference between the knot locations, covariates accounting for moderate growth factors in both clusters, and the small residual variance. We only kept the replications where the four models converged. 

\subsection{Performance Measures}\label{R:Est}
In this section, we present simulation results in terms of performance measures, including the relative bias, empirical SE, relative RMSE and coverage probability for each parameter of the four proposed models. Generally, all four models can estimate parameters unbiasedly, precisely, and generate target confidence interval coverage. We first calculated each performance metric across $1,000$ replications for each parameter of interest under each condition and then summarized the values of each performance metric across all conditions as the corresponding median and range. The summary is provided in the Online Supplementary Document.

All four models produced unbiased point estimates with small empirical SEs, and the magnitude of the relative bias and empirical SE of each parameter across models were comparable. Specifically, the magnitude of the relative biases of the growth factor means, (unexplained) growth factor variances, path coefficients and logistic coefficients were around $0.03$, $0.07$, $0.13$ and $0.10$, respectively. The magnitude of the empirical SE of all parameters except intercept parameters (including intercept means, variances, and path coefficients) was under $0.75$ (i.e., the variances of estimate were under $0.56$). The empirical SE of $\mu_{\eta0}^{(k)}$ and $\psi_{00}^{(k)}$ were around $0.50$ and $2.50$, respectively.

Moreover, four models can estimate parameters accurately, quantified by the relative RMSE that evaluates model performance holistically with consideration of bias and precision. The magnitude of the relative RMSEs of the means and (unexplained) variances of the growth factors were under $0.04$ and $0.33$, respectively. The relative RMSE magnitude of the path and logistic coefficients were around $0.40$ and $0.30$, respectively. 

Overall, the proposed models performed well regarding empirical coverage under most conditions as the median values of coverage probabilities of the mean vector and (unexplained) variance-covariance matrix of growth factors, the path and logistic coefficients were around $0.90$. We noted that the knots' coverage probabilities could be unsatisfactory. We then plotted the coverage probabilities of the class-specific knots stratified by the separation between clusters in Figure \ref{fig:KnotCP} to investigate the pattern. We observed that the coverage probabilities were still around $0.95$ when the separation between clusters was large, although this metric was conservative under other conditions.

\figurehere{2}

\subsection{Clustering Effects}\label{R:Clus}
We assess the clustering effects across all conditions listed in Table \ref{tbl:simu} in this section. For each model under each condition, we first calculated the mean values of accuracy and entropy across $1,000$ replications. We plotted these values stratified by the separation between latent classes, as shown in Figures \ref{fig:Accuracy} and \ref{fig:Entropy}. The mean values of accuracy and entropy were the highest under the conditions with the large difference in knot locations (i.e., $2.0$), followed by those with the medium difference (i.e., $1.5$) and then the small difference (i.e., $1.0$). Specifically, when the difference in knot locations was set as $2$, the ranges of the mean values of accuracy were ($0.78$, $0.91$), ($0.80$, $0.92$), ($0.77$, $0.91$) and ($0.80$, $0.92$) for the FMM, the CP-, GP- and full mixture model, respectively. We also noticed that, on average, the values of accuracy and entropy under conditions with the unbalanced allocation were relatively larger.

\figurehere{3}

\figurehere{4}

In general, the clustering effects of the full mixture model and CP-mixture model were better than the other two models, suggesting that correctly adding covariates with indirect effects helps separate heterogeneous samples, which aligns with the recommendation by existing studies such as \citet{Lubke2007entropy}. We also noticed that including covariates with direct effects only affected accuracy slightly in this simulation study. It is not surprising since one essential factor that affects a clustering algorithm is the separation between class-specific density functions: the larger the separation between two densities is, the easier it is to tell them apart. For a full or GP-mixture model, the class-specific density function is defined by the outcome variable $\boldsymbol{y}$ and the covariates with direct effects $\boldsymbol{x}_{e}$. In the simulation study, the four models share the same class-specific distribution of the outcome $\boldsymbol{y}$ since we fit the four models on the same generated data set. Additionally, we generated standardized covariates $\boldsymbol{x}_{e}$ for each latent class, with which the separation between the class-specific density functions was only affected slightly no matter how large the effects (i.e., the path coefficients) were.

\subsection{Misspecified Model}\label{r:mis}
We first examined the convergence rate of the misspecified CP-mixture model that specifies all four covariates in the multinomial functions. For each condition, we fit the misspecified model on each replication where all four correctly-specified models converged. The convergence rate can achieve at least $92.0\%$, suggesting that the misspecified model produced $920$ replications with a convergent solution out of $1,000$ repetitions, which was still satisfied. However, the misspecified model's performance measures and accuracy were worse than the correctly-specified models' corresponding values. The relative bias and empirical SE of the misspecified model are also provided in the Online Supplementary Document. The estimates from the misspecified model exhibited some bias greater than $10\%$: the relative bias magnitude of the growth factor means, growth factor variances, and logistic coefficients achieved $0.26$, $0.26$, and $0.83$, which were much worse than the corresponding values from the correctly-specified models.

We also plotted the mean values of accuracy and entropy of the misspecified model in Figures \ref{fig:Accuracy} and \ref{fig:Entropy}, respectively. We observed that the mean value of accuracy of the misspecified model could be as low as around $50\%$ (i.e., the probability of having a correct label from a guess). The common conditions that generated such low accuracy values were the small difference in knot locations (i.e., $1.0$), and covariates account for substantial growth factor variances (i.e., $26\%$) at least in one latent class. We noted that the entropy of the misspecified model could be higher than that of the correctly-specified models. This finding suggests that entropy obtained from a mixture model with covariates is no longer a good indicator of the class separation of trajectories alone, which aligns with the finding in an existing study \citep{Stegmann2018Entropy}.

Earlier studies, for example, \citet{Vermunt2010three, Bakk2017two}, have suggested that adding or removing covariates with indirect effects may affect the posterior probabilities, and in turn, the component membership. This also explains why the entropy of mixture models is sensitive to the inclusion of the covariates, especially as the separation between two clusters decreased and when the misspecified covariates accounted for greater within-class variability in the current study. As shown in Equation (\ref{eq:ent}), entropy is a measure based on the average posterior probabilities, and the posterior probabilities are sensitive to the inclusion of covariates, as shown in Equations (\ref{eq:ent_full})-(\ref{eq:ent_FMM}). When the between-class differences quantified by the separation in knot locations decreased, the mixing proportions $g(.)$ dominate the mixture model, and entropy was more indicative of the latent classes forming the covariates than those forming the trajectories. Entropy is a misleading indicator of the class separation of trajectories when the misspecified covariates have higher direct effects on the trajectories.

\subsection{Comparison Among Models}
We compared the four correctly-specified models among themselves and to the misspecified model. We want to point out that the likelihood-based criteria, such as AIC and BIC, cannot be employed to select a model among the four mixture models as the difference in these criteria is mainly from the discrepancy in the estimated likelihood, which is due to their different model structures. In general, the models without covariates with direct effects (i.e., FMM and CP-mixture model) had a much larger estimated likelihood than the other two. We provide a plot with mean values of estimated likelihood, AIC, and BIC across all conditions for the models in the Online Supplementary Document.

We then calculated Dumenci's Latent Kappa to evaluate the agreement between the trajectory clusters obtained from the FMM and those from each of the mixture models with covariates. For each condition, we first calculated the mean value of Dumenci's Latent Kappa across $1,000$ replications and then plotted these values stratified by the separation between clusters in Figure \ref{fig:Agree}. The agreement between trajectory types from two correctly-specified models was high, although it is not exactly perfect (i.e., the Kappa statistic was $1$ as \citet[Chapter~11]{Agresti2012CDA}). Under the conditions with the large separation in the latent classes, the mean values of Dumenci's Latent Kappa were above $0.80$, indicating an almost perfect agreement \citep{Landis1977kappa, Nakazawa2019fmsb} between trajectory clusters from the FMM and those from each of the other three correctly-specified models. However, Dumenci's Latent Kappa between trajectory types from the FMM and those from the misspecified model, which could be below $0.2$ under some conditions, was much lower, suggesting only `slight' agreement \citep{Landis1977kappa, Nakazawa2019fmsb}.  

\figurehere{5}

\section{Employing SEM Forests to Identify Important Covariates}\label{approach}
In this section, we propose to use SEM Forests to shrink the covariate space when building a mixture model. As shown in Table \ref{tbl:approach}, we examined eight scenarios with the different relative importance of direct effects against indirect effects. We considered one cluster in the first three scenarios and two latent classes in the other five scenarios. For the scenarios with one cluster, we set the covariates to account for $2\%$ (small), $13\%$ (moderate) and $26\%$ (substantial) variability of growth factors, respectively. For the other scenarios, similar to the simulation design, we fixed the logistic coefficients and varied path coefficients to adjust the relative importance. We also standardized all covariates. 

\tablehere{3}

Figures \ref{fig:Step1} and \ref{fig:Step2} are the general steps to examine the variable importance of the scenarios with one and multiple cluster(s), respectively. We first generated data and constructed a latent growth curve model in the \textit{R} package \textit{OpenMx} and this one-group model serves as a template model for this generated data set \citep{Brandmaier2016semForest}. As shown in these figures, the input of the SEM Forests algorithm, which is available in the \textit{R} package \textit{semtree} \citep{User2020semtree}, is the template model, the original data set, and the pool of candidate covariates. One output of SEM Forests is the variable importance in terms of predicting the model-implied mean vector and variance-covariance structure. Note that we added two noise variables whose importance is supposed to be zero when building forests. In this study, the tree parameter setting is `bootstrap' as the sampling method, $128$ trees\footnote{We decided to use $128$ trees as \citet{Oshiro2012Notree} showed that from $128$ trees, there are no more significant difference between the forest with $256$, $512$, $1024$, $2048$ and $4096$ trees by analyzing $29$ real-world data sets.}, and $2$ subsampled covariates at each node\footnote{As \citet{Brandmaier2016semForest}, suppose there are $m$ potential predictors, the size of the set of candidate predictors, $c$, at each node could be set as either $1$, $2$,  $c=\log_{2}(m)+1$, $c=\sqrt{m}$, or $c=m/3$. In this section, we consider $c=m/3$ and select $c=2$ as $m=6$.}. We provide a demonstration on building a SEM Forests model in the online appendix (\url{https://github.com/Veronica0206/Extension_projects}).

\figurehere{6}

Figure \ref{fig:varImp} plots variable importance for each scenario that we considered in Table \ref{tbl:approach}. For the conditions that only include covariates with direct effects and noise variables (i.e., Figure \ref{fig:varImp}a, Figure \ref{fig:varImp}b and Figure \ref{fig:varImp}c), the algorithm worked well to distinguish the covariates from the noise variables and weighted more on the covariates when they account for more variability of growth factors. For the conditions including the two types of covariates with noise (i.e., Figure \ref{fig:varImp}d-Figure \ref{fig:varImp}h), the SEM Forests model performed well to distinguish the signal (i.e., the covariates that have effects) from the noise, and the rank of importance scores of the covariates may change with the relative importance of the direct effects against the indirect effects. 

\figurehere{7}

Based on the patterns demonstrated in Figure \ref{fig:varImp}, we propose a possible approach to identify covariates for a mixture model in the SEM framework: (1) for a given data set, fit a template model in the \textit{R} package \textit{OpenMx}, (2) regress the template model on a pool of candidate covariates to build a SEM Forests model using the \textit{R} package \textit{semtree} and obtain variable importance scores, (3) decide a threshold in terms of `importance' and select covariates with high importance. 

\section{Application}\label{application}
We have three goals in the application section. The first goal is to demonstrate how to employ the SEM Forests to identify the covariates with high importance and then investigate the direct and indirect effects that the covariates have on the heterogeneity in trajectories by employing the proposed models. The second goal is to demonstrate how to extend the proposed model by allowing the covariates with direct effects to be time-varying. We then conduct sensitivity analyses, where we keep the functional form of the within-class model but use stepwise procedures to add covariates with indirect effects. A random sample with $500$ students was selected from the Early Childhood Longitudinal Study Kindergarten Cohort: 2010-11 (ECLS-K: 2011) with complete records of repeated mathematics achievement scores, demographic information (sex, race/ethnicity, and age in months at each wave), baseline socioeconomic status (family income and the highest education level between parents), baseline school information (school type and location), repeated measurements of teacher-reported approach-to-learning, baseline teacher-reported social skills (including self-control ability, interpersonal skills, externalizing problem and internalizing problem), and baseline teacher-reported children behavior question (including attentional-focus and inhibitory-control).

ECLS-K: $2011$ is a nationwide representative longitudinal sample consists of US children from around $900$ kindergarten programs that started from $2010-2011$ school year. Student's mathematics IRT scores were evaluated in nine waves: the fall and spring semester of kindergarten, first and second grade, respectively, as well as the spring semester of $3^{rd}$, $4^{th}$ and $5^{th}$ grade, respectively. As \citet{Le2011ECLS}, this study only sampled around $30\%$ students in the fall semester of $2011$ and $2012$. We employed children's age (in months) to obtain individual measurement occasions. The selected sample ($n=500$) consists of $53.0\%$ boys and $47.0\%$ girls. Additionally, the sample was represented by White ($52.4\%$), Black ($3.2\%$), Latinx ($29.6\%$), Asian ($8.8\%$) and others ($6.0\%$). We then dichotomized the variable race/ethnicity to be White ($52.4\%$) and others ($47.6\%$). The highest parents' education (ranged from $0$ to $8$) and family income (ranged from $1$ to $18$) were treated as continuous variables for this analysis, and the corresponding mean (SD) was $5.34$ ($1.93$) and $12.09$ ($5.31$), respectively.

\subsection{Main Analysis}\label{app:main}
We first fit a bilinear spline growth curve model for these repeated mathematics scores and built a SEM Forests model to identify the covariates with the greatest variable importance. As shown in Figure \ref{fig:MathvarImp}, the covariates with the highest variable importance scores were family income ($113.04$) and parents' highest education level ($110.52$),  followed by approach-to-learning ($55.73$) and attentional-focus ($43.41$). We decided to keep these four covariates with sex and race/ethnicity to build the following six models:
\begin{enumerate}
\item {Model $1$: A growth mixture model without any covariates,}
\item {Model $2$: A CP-mixture model with family income, parents' highest education, sex and race/ethnicity in the multinomial functions,}
\item {Model $3$: A CP-mixture model with all six covariates in the multinomial functions,}
\item {Model $4$: A GP-mixture model with attentional-focus and approach-to-learning in all within-class models,}
\item {Model $5$: A GP-mixture model with family income and parents' highest education in all within-class models,}
\item {Model $6$: A full mixture with attentional-focus and approach-to-learning in all within-class models and the other four covariates in the multinomial functions.}
\end{enumerate}

\figurehere{8}

Following the convention in the SEM literature, we decided the number of latent classes without any covariates. We fit a bilinear spline growth model and bilinear growth mixture models with two, three, and four classes. Information criteria such as the BIC suggested that the model with three latent classes was the `best' among the four candidate models. Model 1, the growth mixture model, served as a reference model in this section. We evaluate the agreement, quantified by Dumenci's Latent Kappa \citep{Dumenci2011kappa}, between the trajectory clusters obtained from the growth mixture model and each of the other five models. We provide the membership agreement in Table \ref{tbl:compare}. We also include the CPU time charged for the execution of each model in Table \ref{tbl:compare} given that the computational budget is an important consideration in practice. 

\tablehere{4}

\subsubsection*{Growth Mixture Model}
Table \ref{tbl:FMM_est} summarizes the estimates of the class-specific growth factors. Based on the estimates, we obtained the model implied trajectory of each latent class, as shown in Figure \ref{fig:FMM_traj}. The estimated proportions in Class $1$, $2$ and $3$ were $17.20\%$, $45.00\%$ and $37.80\%$, respectively. Post-knot development in mathematics skills in three classes slowed down substantially. The transition to the slower growth rate occurred at $82$, $109$ and $98$ months in Class $1$, $2$ and $3$, respectively. The students grouped into Class $1$ had the lowest levels of mathematics achievement in general (the estimated fixed effects of the intercept and two slopes were $24.046$, $1.921$ and $0.925$ per month). The second cluster's initial status and developmental rates were lower than those of the first cluster; however, students in Class $2$ had better performance in mathematics as their transition to the slower growth rate occurred $1.5$ years later than students in Class $1$. Students in Class $3$ had the best performance in mathematics throughout the study duration. 

\tablehere{5}

\subsubsection*{Cluster Predictor Mixture Model}
We built two CP-mixture models, Model $2$ and Model $3$. We included socioeconomic status (i.e., family income and parents' highest education) and demographic information (i.e., sex and race) in the multinomial functions of Model 2. As shown in Figure \ref{fig:Gating_traj} and Table \ref{tbl:Gating_est}, the estimated proportions and predicted trajectories slightly changed when adding these selected covariates. On further investigation, $53$ out of $500$ students were assigned to a different group by Model 1 and Model 2. The Dumenci's Latent Kappa between student clusters from two models was $0.83$ with $95\%$ CI ($0.79$, $0.88$), suggesting an almost perfect agreement \citep{Landis1977kappa, Nakazawa2019fmsb}. From this CP-mixture model, we obtained the covariates' effects, as shown in Table \ref{tbl:Gating_est}. Specifically, with all other covariates, boys are more likely to be in Class $2$ and Class $3$. Higher parents' education increased the probability of being in Class $3$.

\tablehere{6}

In Model $3$, the other CP-mixture model, we included all six covariates in the multinomial functions. As shown in Figure \ref{fig:misGating_traj}, the estimated proportions and predicted trajectories also only slightly changed from those obtained from the finite mixture model, and the Dumenci's Latent Kappa was $0.84$ with $95\%$ CI: ($0.80$, $0.88$). The effects on Class $2$ and Class $3$ of family income, parents' highest education, sex and race/ethnicity of two CP-mixture models were the same (in terms of effect size and direction). Additionally, higher attentional-focus and a better approach-to-learning increased the likelihood of having better mathematics achievement. 

\subsubsection*{Growth Predictor Mixture Model}
Next, we constructed two GP-mixture models with the BLSGM-TICs as the within-class model, Model $4$ and Model $5$. In Model $4$, we included the variables approach-to-learning and attentional-focus to explain the within-cluster trajectory heterogeneity. As shown in Figure \ref{fig:Expert_traj} and Table \ref{tbl:Expert_est}, the estimated proportions and predicted trajectories slightly changed from those obtained from the growth mixture model, and the Dumenci's Latent Kappa was $0.85$ with $95\%$ CI: ($0.81$, $0.89$).

An important piece of information obtained from the GP-mixture model is the estimates of the covariates with direct effects, as shown in Table \ref{tbl:Expert_est}. First, the estimated means of the standardized attentional-focus/approach-to-learning was negative ($-0.513$/$-0.446$), around zero ($-0.025$/$-0.052$) and positive ($0.240$/$0.249$) for Class $1$, Class $2$ and Class $3$, respectively. It suggests that the region with a higher value of attentional-focus and approach-to-learning was associated with higher mathematics scores and \textit{vice versa}. Additionally, the effects of those covariates varied across classes, although such effects were positive in general. 

\tablehere{7}

In Model $5$, we put the covariates family income and parents' highest education to explain the within-cluster trajectory heterogeneity. From Figure \ref{fig:misExpert_traj}, the estimated proportions and predicted trajectories changed a lot from those of Model $1$, the growth mixture model. The Dumenci's Latent Kappa was $0.11$ with $95\%$ CI: ($0.05$, $0.17$), suggesting slight agreement. Upon further examination, the estimated means of the standardized family income (parents' highest education level) was $-0.826$ ($-0.612$), $0.869$ ($0.601$) and $-0.304$ ($-0.128$) for Class $1$, Class $2$ and Class $3$, respectively. This nonlinear relationship between socioeconomic status and academic performance is one possible explanation for this low Dumenci's Latent Kappa.

\subsubsection*{Full Mixture Model}
Finally, we built Model $6$, a full mixture model, with the variables attentional-focus and approach-to-learning in each cluster and the other four in the multinomial functions. Figure \ref{fig:Full_traj} and Table \ref{tbl:full_est} suggest that the estimated trajectories and proportions were similar to those produced by the GMM. The Dumenci's Latent Kappa between the student clusters from the GMM and those from the full mixture was $0.87$ with $95\%$ CI ($0.83$, $0.91$). The direct and indirect effects on the heterogeneity in the trajectories were similar to those in Model $4$ and Model $2$, respectively.

\tablehere{8}

\subsection{Possible Extension of the Proposed Mixture Model}
In this section, we demonstrate how to extend the proposed mixture model in the SEM framework by allowing for a time-varying covariate, approach-to-learning, in each latent class. We included sex, race/ethnicity, family income, and parents' highest education in the multinomial functions. We provide the estimates of the extended model in the Online Supplementary Document. Although the estimated change patterns and proportions were similar to those of the models in Section of Main Analysis, class membership showed substantial change. Dumenci's Latent Kappa between the trajectory types from the GMM and those from the extended model was $0.28$ with $95\%$ CI ($0.22$, $0.35$). 

Similar to the GP-mixture model (Model $4$) and the full mixture model (Model $6$), the estimated means of the standardized approach-to-learning were negative (about $-1$), around $0$, and positive (about $1$) in Class $1$, $2$ and $3$, respectively over time. It suggests the time-varying covariate at each wave was split into three segments to associate with mathematics trajectories in each latent class, similar to the covariates with direct effects in Model $4$ and Model $6$. Additionally, the effects on the heterogeneity in the trajectories of the time-varying covariate varied across latent classes. We also noted that girls are more likely to be in Class $2$ and Class $3$ when the within-class model was set as a BLSGM with the time-varying covariate, which is different from the effect of the variable sex on the class-formation in Model $2$, Model $3$ and Model $6$. One possible explanation for this opposite effect of the covariate sex is the difference in the membership discussed above. 

\subsection{Sensitivity Analysis}
In this section, we built stepwise mixture models as a sensitivity analysis. We consider two types of stepwise approaches, the standard three-step method \citep{Clogg1995three} and the two-step method \citep{Bakk2017two}. We decided to fit a standard three-step method with modal assignment as it is one of the most popular methods in psychology \citep{Hsiao2020mediation}. We chose the two-step approach as \citet{Bakk2017two} has shown that this method is a competitive alternative to the three-step maximum likelihood approach \citep{Vermunt2010three} by simulation studies.

The first step of either the two- or three-step approach is building a GP-mixture model (i.e., Model $4$ in Section of Main Analysis). As shown in the simulation studies and the empirical example analyses, the estimated proportions and within-class parameters from the GP-mixture model and the full mixture model (i.e., one-step method) were very similar. For the standard three-step method, we assigned each student to the cluster with the highest probability and then constructed a multinomial logistic regression by the \textit{R} package \textit{nnet} \citep{Venables2002Statistics}. We estimated logistic coefficients for the two-step approach by fixing the within-class parameters as their estimates from the Model $4$. We provide the estimates of logistic coefficients of two stepwise methods in the Online Supplementary Document.

Generally, the estimates of logistic coefficients of continuous covariates of both stepwise methods were very similar to those provided by the full mixture model, although the estimated coefficients of binary covariates were different. Specifically, both the point estimates and the standard errors were larger in the stepwise models. 

\section{Discussion}\label{discussion}
This article extends MoE models to the SEM framework. The full mixture model and its three possible reduced models are multivariate methods designed to uncover sample heterogeneity underlying longitudinal or cross-sectional data sets. We linked these four models to the corresponding counterpart in the SEM framework. The FMMs and CP-mixture models have received considerable attention in the SEM literature. On the contrary, the GP-mixture models and full mixture models are relatively novel and have only been implemented in limited settings. 

For the full mixture model and its three possible reduced versions, we performed in-depth investigations in terms of convergence rate, performance measures (including relative bias, empirical SE, relative RMSE and coverage probability) and clustering effects (including accuracy and entropy) by simulation studies. We also illustrated the proposed models using a real-world data set from a longitudinal study of mathematics ability. The results demonstrate the proposed models' valuable capability to identify latent classes and examine direct and indirect effects on the sample heterogeneity of covariates. The proposed model can be further extended in the SEM framework. For example, within-class covariates can also be time-varying. We demonstrate how to implement and interpret the extended model in the Application section. 

We construct the proposed model using the one-step approach in a stepwise fashion. On the one hand, multiple existing simulation studies have demonstrated that the one-step approach outperforms the two-step or three-step approach concerning performance metrics such as bias and RMSE. On the other hand, to address the major critiques on the one-step method, the number of clusters and the covariates added in the model need to be decided before constructing the proposed models. As the current recommended approach in the SEM literature, we decided the number of latent classes without covariates. We also propose to shrink covariate space by leveraging the SEM Forests. 

We conducted a sensitivity analysis in the empirical example to examine whether the two-step or three-step approach would provide different insights from the one-step method. As shown in the Application section, the proposed model can also be constructed by stepwise approaches. The estimates of logistic coefficients of continuous covariates were very similar to those from the full mixture model, although the estimated coefficients of binary covariates were different. 

\subsection{Practical Considerations}
The current study focuses on applying mixture models to analyze the heterogeneity in nonlinear trajectories and the impact that covariates have on such heterogeneity. As demonstrated in the Application section, the recommended steps are:
\begin{enumerate}
\item {Without any covariates, build a latent growth curve model and growth mixture models with different numbers of clusters to select the optimal number by the BIC,}
\item {Construct a SEM Forest model using the latent growth curve model from (1) (the template model) and the original data set with a pool of candidate covariates to select covariates with high importance,}
\item {Among the covariates from (2), decide which covariates have direct effects and include them in the latent classes and the others in the multinomial functions to construct the proposed model(s).}
\end{enumerate}

Other than the stepwise list, we also want to provide a set of recommendations for possible issues that empirical researchers may face in practice. First, it is not our aim to show that one model is universally preferred. We recommend selecting the model according to domain knowledge and research interests. We decided to add four covariates with the highest importance and demographic information in the model. In the full mixture model, we assume that the variables teacher-reported attentional-focus and approach-to-learning affect the trajectory heterogeneity directly as they reflect students' ability and potential. In contrast, other covariates are more about the environment that students live in; accordingly, we placed them in the multinomial functions assuming that these covariates have indirect effects. In addition, as discussed in Section Comparison Among Models, the likelihood-based criteria, such as the AIC and BIC, cannot be used to select among these four mixture models due to their different model structures.

As shown in Section Misspecified Model, entropy does not help select a model among the full mixture model and its three reduced versions in an exploratory study. It is not our goal to explore this metric all-inclusively in this current project; still, we want to add a note for empirical researchers. Entropy itself is still a good metric to tell class separation. However, entropy obtained from a growth mixture model with covariates only reflects the separation based on the trajectories and covariates. Although our simulation results suggest that entropy is a good indicator of the fraction of all correctly classified trajectories if we specified covariates correctly, we do not recommend using entropy from a mixture model with covariates as an indicator of the class separation of trajectories alone as it is impossible to know if a covariate is specified correctly in practice. We have shown the discrepancy between the patterns of accuracy (i.e., the fraction of all correctly classified trajectories) and entropy from the misspecified model.

Second, in an exploratory study, researchers may hope that including covariates does not substantially influence class membership. To evaluate whether adding covariates affects trajectory clusters, we recommend assessing the agreement between the trajectory types obtained from the FMM and those from a mixture model with covariates, which can be quantified by the Dumenci's Latent Kappa \citep{Dumenci2011kappa}. Adding covariates can also be driven by a specific research question. For example, we may build Model $5$ if the research question is to group the relationship between mathematics trajectories and socioeconomic status, although assuming the socioeconomic status can directly affect the trajectory heterogeneity changed the membership heavily. Similarly, the extended model with the time-varying covariate allows us to assess the effect of approach-to-learning on mathematics development over time in each latent class by grouping the association between the mathematics trajectories and the repeated measures of the covariate.

Additionally, we propose one possible approach to select covariates by leveraging the SEM Forests algorithm. As demonstrated in Section of Employing SEM Forests to Identify Important Covariates, SEM Forests identified covariates with high importance efficiently. Although it is not our aim to comprehensively investigate the SEM Forests, we still want to add two notes about model construction and variable importance for empirical researchers. First, when several covariates are highly correlated, a covariate with a relatively low importance score is not necessarily unimportant. For example, the correlation between attentional-focus and inhibitory-control is $0.7$, yet the variable inhibitory-control has a lower importance score than attentional-focus. One more reasonable explanation for this phenomenon is that the SEM Forests model can address the collinearity between covariates to some extend. Imagine that the algorithm selects the attentional-focus to predict the model-implied means and variance-covariance structure. This covariate also takes over the predicting responsibility belonged to the variable inhibitory-control due to the high correlation between these two variables. The algorithm will not consider splitting on the variable inhibitory-control. Second, the SEM Forests model has several hyperparameters, including $c$ (the size of the set of candidate predictors at each node) and the number of trees. Researchers can also select a sampling method between `subsample' and `bootstrap'. In this article, we set the hyperparameters following \citet{Brandmaier2016semForest} and the machine learning literature. We selected `bootstrap' as the sampling method to follow the convention in the RFs model. However, if building a forest is the aim, these hyperparameters needed to be tuned \citep{Brandmaier2016semForest}. 

Moreover, the selection of starting values of parameters is important when building a complicated model for a real-world data analysis. A set of proper starting values helps improve the likelihood of convergence and alleviate the computational burden. Empirically, we select proper starting values by (1) deciding appropriate scales for parameters and then (2) searching starting values around the scales from which a model can achieve convergent status (i.e., the status code $0$ in the \textit{OpenMx}). We recommend plotting the trajectories to obtain the appropriate scales of growth factor means. Sometimes, it may not be helpful to use the same set of starting values across clusters. We then recommend building a parsimonious model (i.e., a FMM with fixed variance-covariance matrices of growth factors) to estimate the means of class-specific growth factors that can serve as the starting values. An option for the scale of variances and covariances of growth factors is $1$ and $0$, respectively. The selection of scales of the $\boldsymbol{x}_{e}$-related parameters is relatively straightforward as we standardize $\boldsymbol{x}_{e}$'s. Additionally, we set the scale of path coefficients and logistic coefficients as $0.5$ and $1.0$, respectively. After obtaining appropriate scales, we do not recommend employing a grid search\footnote{In a grid search, we create a list of starting values for each parameter, run the model with different combinations of the values across all parameters, and see which set could lead to a convergence solution.} to tune the starting values given a large number of parameters. Instead, we suggest utilizing the function \textit{mxTryHard()} in the \textit{R} package \textit{OpenMx} that allows for multiple attempts to fit a model until generating a convergent solution or reaching the specified maximum number of runs. At each attempt, each starting value is multiplied by a random draw\footnote{For replication purposes, we recommend setting and recording seeds for random sampling.} from a pre-specified distribution. 

Although the purpose of our examples was pedagogical, in practice, researchers should be aware of the multiple statistical tests conducted when fitting a growth mixture model with covariates and should consider adjusting the tests to control the type I error or false discovery rate. Furthermore, decisions about the importance of a finding should not be based on p-values alone, but also consider effect sizes, prior evidence, and alternative explanations \citep{Wasserstein2019Pvalue}.

Last, in the Application section, it took $8$, $12$, $38$ and $39$ minutes to fit Model $1$ (a FMM), Model $2$ (a CP-mixture model), Model $4$ (a GP-mixture model), and Model $6$ (a full mixture model), respectively. Note that the CPU time can be reduced to about $30\%$ if a structured time matrix is used rather than the individual measurement occasions.

\subsection{Methodological Considerations and Future Directions}
This study has shown multiple directions in need of further investigation. First, as stated earlier, we assume that one covariate does not have direct and indirect effects on sample heterogeneity simultaneously based on conceptual and technical considerations. We included the covariates attentional-focus and approach-to-learning, which are assumed to have direct effects on the trajectory heterogeneity, in the within-class model. The clustering algorithm also divides them into regions with negative, around zero and positive mean values when separating trajectories into clusters with lower, medium and higher mathematics achievement (Models $4$ and $6$). Conceptually, it may not be reasonable to include the two covariates in the multinomial functions as the predictors of latent classes with the two covariates' class-specific mean values. To further demonstrate our points, suppose we include the two covariates as the predictors of latent classes. Since the covariates have been divided, the increased odds of being in Class $2$ or $3$ compared to Class $1$ that associated with a one SD difference in these predictors when the covariates are negative could be different from the values when the covariates are positive (i.e., the linear assumption between the log-odds and the covariates could be violated technically).

However, a covariate that has direct effects on sample heterogeneity often has indirect effects. Based on the class-specific estimated means of the standardized path covariates (Models $4$ and $6$), a more proper gating function is to separate the covariate space into multiple regions (for example, a negative, around zero, and a positive region). The estimates of such gating functions would be the `boundary' of each region. This issue has been addressed in the machine learning literature. \citet{Jordan1993MoE} have developed a much more flexible model by allowing for a multilevel gating function to give the hierarchical mixture-of-experts (HMoE) model. Similar to a tree-based algorithm, the HMoE model needs to select a covariate to split and the threshold value at each level of gating functions. Then a within-class model is constructed in each region of the covariate space. Additionally, for the mixture model, the model fitting is just the first part of the story as currently, the model only allows for (generalized) linear relationships between covariates and class membership or growth factors in the SEM framework. Model diagnostics for evaluating such linear assumptions can help strengthen the story. It is worth conducting further simulation studies to evaluate and develop at least visualization tools for model diagnostics.

Second, the impact of a covariate with direct effects on the clustering algorithm depends on its influence on the separation between latent classes. If we generated $\boldsymbol{x}_{e}$ from the same distribution across latent classes so that the separation between clusters did not change, as shown in the simulation study, $\boldsymbol{x}_{e}$ only affected clustering results slightly. On the contrary, we learned that adding a covariate with direct effects may affect trajectory clusters meaningfully (Model $5$ and the extended model). One reasonable explanation is that $\boldsymbol{x}_{e}$ affects the separation between latent classes since the estimates of the class-specific means of the $\boldsymbol{x}_{e}$ varied a lot across clusters. It is worth conducting a further simulation study that allows for varying parameters of $\boldsymbol{x}_{e}$ to investigate the conditions under which and how those covariates influence the clustering algorithm. 

Third, in the Application section, we demonstrate how to extend the proposed mixture model by allowing for time-varying covariates. Its performance, such as estimates and clustering effects, needs to be evaluated by a simulation study. Additionally, in the sensitivity analyses, we build the proposed mixture model using stepwise approaches. A further simulation study is needed to compare these stepwise approaches to the proposed full mixture model, especially for the logistic coefficients of binary covariates. 

Last, the project proposes leveraging SEM Forests to select the covariates with high effects on sample heterogeneity and then building models to evaluate their effect sizes. We tested SEM Forests' performance using generated data sets under several scenarios with different weights on covariates with direct and indirect effects along with noise variables. As only a few articles other than the original article detailing this algorithm \citep{Brandmaier2016semForest}, it is worth conducting more comprehensive simulation studies to assess its performance. It is also worth examining how to tune the hyperparameters, such as the number of trees and the size of covariates selected at each node.

\subsection{Concluding Remarks}
In this article, we extend the MoE model, which allows for examining direct and indirect effects of covariates simultaneously, to the SEM framework to investigate the heterogeneity in nonlinear trajectories. Overall, we have shown the performance and application of MoE models using bilinear spline functional form with an unknown knot and covariates as the within-class model. Note that the nonlinear underlying change patterns could be other functional forms. We provide two common parametric functional forms, quadratic and Jenss-Bayley, in the online appendix for the researchers who are willing to employ them. 

\bibliographystyle{apalike}
\bibliography{Extension2}

\appendix
\renewcommand{\theequation}{A.\arabic{equation}}
\setcounter{equation}{0}

\renewcommand{\thesection}{Appendix \Alph{section}}
\renewcommand{\thesubsection}{A.\arabic{subsection}}
\setcounter{subsection}{0}
\section{Formula Derivation}
\subsection{Reparameterization of Class-specific Growth Factors}\label{Supp:1A}
In the original setting of a bilinear spline growth model, we have three growth factors for each individual to define the underlying functional form of repeated measures: the measurement at $t_{0}$ ($\eta_{0i}$) and one slope of each stage ($\eta_{1i}$ and $\eta_{2i}$, respectively). To estimate the knot in each latent class, we need to reparameterize these growth factors to be the measurement at the knot ($\eta_{0i}+\eta_{1i}\gamma$), the mean of two slopes ($\frac{\eta_{1i}+\eta_{2i}}{2}$), and the half difference between two slopes ($\frac{\eta_{2i}-\eta_{1i}}{2}$) for the $i^{th}$ individual \citep[Chapter~9]{Seber2003nonlinear}.

\figurehere{A.1}

\citet{Tishler1981nonlinear} and \citet{Seber2003nonlinear} have proved that a linear-linear regression model can be expressed as either the maximum or minimum response value of two trajectories. \citet{Liu2019knot} and \citet{Liu2019BLSGM} extended such expressions to the framework of BLSGM and showed that two possible forms of bilinear spline for the $i^{th}$ individual as such in Figure \ref{fig:E2_2cases}. In the left panel ($\eta_{1i}>\eta_{2i}$), the measurement $y_{ij}$ is always the minimum value of two lines and $y_{ij}=\min{(\eta_{0i}+\eta_{1i}t_{ij}, \eta_{02i}+\eta_{2i}t_{ij})}$. The measurements pre- and post-knot can be unified
\begin{equation}\label{eq:left}
\begin{aligned}
y_{ij} &= \min{(\eta_{0i} + \eta_{1i}t_{ij}, \eta_{02i} + \eta_{2i}t_{ij})}\\
&= \frac{1}{2}\big(\eta_{0i} + \eta_{1i}t_{ij} + \eta_{02i} + \eta_{2i}t_{ij} - 
|\eta_{0i} + \eta_{1i}t_{ij} - \eta_{02i} - \eta_{2i}t_{ij}|\big)\\
&= \frac{1}{2}\big(\eta_{0i} + \eta_{1i}t_{ij} + \eta_{02i} + \eta_{2i}t_{ij}\big) - 
\frac{1}{2}\big(|\eta_{0i} + \eta_{1i}t_{ij} - \eta_{02i} - \eta_{2i}t_{ij}|\big)\\
&= \frac{1}{2}\big(\eta_{0i} + \eta_{02i} + \eta_{1i}t_{ij} + \eta_{2i}t_{ij}\big) - 
\frac{1}{2}\big(\eta_{1i} - \eta_{2i}\big)|t_{ij} - \gamma|\\
&= \eta_{0i} + \eta_{1i}\big(t_{ij}-\gamma\big) + \eta_{2i}|t_{ij} - \gamma|\\
&= \eta_{0i} + \eta_{1i}\big(t_{ij}-\gamma\big) + \eta_{2i}\sqrt{(t_{ij} - \gamma)^2},
\end{aligned}
\end{equation}
where $\eta_{0i}$, $\eta_{1i}$ and $\eta_{2i}$ are the measurement at the knot, the mean of two slopes, and the half difference between two slopes of the trajectory of $y_{ij}$. With straightforward algebra, the outcome $y_{ij}$ of the bilinear spline in the right panel, where the measurement $y_{ij}$ is always the maximum value of two lines, has the same final expression in Equation \ref{eq:left}. We obtain the class-specific reparameterized growth factors by applying such transformation for three growth factors in each latent class.

\subsection{Class-specific Transformation and Inverse-transformation Matrices}\label{Supp:1B}
Suppose $\boldsymbol{f}: \mathcal{R}^{3}\rightarrow \mathcal{R}^{3}$ is a function, which takes a point $\boldsymbol{\eta}_{i}\in\mathcal{R}^{3}$ as input and produces the vector $\boldsymbol{f}(\boldsymbol{\eta}_{i})\in\mathcal{R}^{3}$ (i.e., $\boldsymbol{\eta}_{i}^{'}\in\mathcal{R}^{3}$) as output. By the multivariate Delta Method\footnote{In this study, both $\boldsymbol{f}$ and $\boldsymbol{h}$ are linear functions. Under this scenario, the mean and variance can be derived using the theorem for calculating the mean and variance of linear combinations. The results obtained by the theorem and the Delta Method are identical.} \cite[Chapter~1]{Lehmann1998Delta}, for the $i^{th}$ individual in the $k^{th}$ cluster
\begin{equation}\label{eq:trans_fun}
\boldsymbol{\eta}_{i}^{'}=\boldsymbol{f}(\boldsymbol{\eta}_{i})\sim N\bigg(\boldsymbol{f}(\boldsymbol{\mu}^{(k)}_{\boldsymbol{\eta}}), \boldsymbol{\nabla_{f}}(\boldsymbol{\mu}^{(k)}_{\boldsymbol{\eta}})\boldsymbol{\Psi}^{(k)}_{\boldsymbol{\eta}}\boldsymbol{\nabla}^{T}_{\boldsymbol{f}}(\boldsymbol{\mu}^{(k)}_{\boldsymbol{\eta}})\bigg), 
\end{equation}
where $\boldsymbol{\mu}^{(k)}_{\boldsymbol{\eta}}$ and $\boldsymbol{\Psi}^{(k)}_{\boldsymbol{\eta}}$ are the mean vector and variance-covariance matrix of class-specific growth factors in the original framework, and $\boldsymbol{f}$ is defined as
\begin{equation}\nonumber
\boldsymbol{f}(\boldsymbol{\eta}_{i})=\left(\begin{array}{rrr}
\eta_{0i}+\gamma^{(k)}\eta_{1i} & \frac{\eta_{1i}+\eta_{2i}}{2} & \frac{\eta_{2i}-\eta_{1i}}{2}
\end{array}\right)^{T}.
\end{equation}

Similarly, suppose $\boldsymbol{h}: \mathcal{R}^{3}\rightarrow \mathcal{R}^{3}$ is a function, which takes a point $\boldsymbol{\eta}_{i}^{'}\in\mathcal{R}^{3}$ as input and produces the vector $\boldsymbol{h}(\boldsymbol{\eta}_{i}^{'})\in\mathcal{R}^{3}$ (i.e., $\boldsymbol{\eta}_{i}\in\mathcal{R}^{3}$) as output. By the multivariate Delta Method, for the $i^{th}$ individual in the $k^{th}$ cluster
\begin{equation}\label{eq:inverse_fun}
\boldsymbol{\eta}_{i}=\boldsymbol{h}(\boldsymbol{\eta}_{i}^{'(k)})\sim N\bigg(\boldsymbol{h}(\boldsymbol{\mu}^{'(k)}_{\boldsymbol{\eta}}), \boldsymbol{\nabla_{h}}(\boldsymbol{\mu_{\eta}^{'(k)}})\boldsymbol{\Psi}^{'(k)}_{\boldsymbol{\eta}}\boldsymbol{\nabla}^{T}_{\boldsymbol{h}}(\boldsymbol{\mu_{\eta}^{'(k)}})\bigg), 
\end{equation}
where $\boldsymbol{\mu}^{'(k)}_{\boldsymbol{\eta}}$ and $\boldsymbol{\Psi}^{'(k)}_{\boldsymbol{\eta}}$ are the mean vector and variance-covariance matrix of class-specific growth factors in the reparameterized framework, and $\boldsymbol{h}$ is defined as
\begin{equation}\nonumber
\boldsymbol{h}(\boldsymbol{\eta}_{i}^{'})=\left(\begin{array}{rrr}
\eta^{'}_{0i}-\gamma^{(k)}\eta^{'}_{1i}+\gamma^{(k)}\eta^{'}_{2i} & \eta^{'}_{1i}-\eta^{'}_{2i} & \eta^{'}_{1i}+\eta^{'}_{2i}
\end{array}\right)^{T}.
\end{equation}

Based on Equations (\ref{eq:trans_fun}) and (\ref{eq:inverse_fun}), the transformation between the mean vector of the class-specific growth factors in the two parameter-spaces can be conducted by $\boldsymbol{\mu}^{'(k)}_{\boldsymbol{\eta}}
=\boldsymbol{f}(\boldsymbol{\mu}^{(k)}_{\boldsymbol{\eta}})$ and $\boldsymbol{\mu}^{(k)}_{\boldsymbol{\eta}}
=\boldsymbol{h}(\boldsymbol{\mu}^{'(k)}_{\boldsymbol{\eta}})$, respectively. We can also define matrices $\boldsymbol{\nabla_{f}}(\boldsymbol{\mu}^{(k)}_{\boldsymbol{\eta}})$ and $\boldsymbol{\nabla_{h}}(\boldsymbol{\mu_{\eta}^{'(k)}})$ to transform the (unexplained) variance-covariance matrix of the class-specific growth factors in the two parameter-spaces as
\begin{equation}\nonumber
\begin{aligned}
\quad\quad\boldsymbol{\Psi}^{'(k)}_{\boldsymbol{\eta}} 
&= \boldsymbol{\nabla_{f}}(\boldsymbol{\mu}^{(k)}_{\boldsymbol{\eta}})\boldsymbol{\Psi}^{(k)}_{\boldsymbol{\eta}}\boldsymbol{\nabla}^{T}_{\boldsymbol{f}}(\boldsymbol{\mu}^{(k)}_{\boldsymbol{\eta}})\\
&=\left(\begin{array}{rrr}
1 & \gamma^{(k)} & 0 \\
0 & 0.5 & 0.5 \\
0 & -0.5 & 0.5 
\end{array}\right)\boldsymbol{\Psi}^{(k)}_{\boldsymbol{\eta}}\left(\begin{array}{rrr}
1 & \gamma^{(k)} & 0 \\
0 & 0.5 & 0.5 \\
0 & -0.5 & 0.5
\end{array}\right)^{T} \ \ \ \ \ \ \ \\
\end{aligned}
\end{equation}
and
\begin{equation}\nonumber
\begin{aligned}
\boldsymbol{\Psi}^{(k)}_{\boldsymbol{\eta}} 
&=\boldsymbol{\nabla_{h}}(\boldsymbol{\mu_{\eta}^{'(k)}})\boldsymbol{\Psi}^{'(k)}_{\boldsymbol{\eta}}\boldsymbol{\nabla}^{T}_{\boldsymbol{h}}(\boldsymbol{\mu_{\eta}^{'(k)}})\\
&=\left(\begin{array}{rrr}
1 & -\gamma^{(k)} & \gamma^{(k)} \\0 & 1 & -1 \\0 & 1 & 1 
\end{array}\right) \boldsymbol{\Psi}^{'(k)}_{\boldsymbol{\eta}}\left(\begin{array}{rrrr}
1 & -\gamma^{(k)} & \gamma^{(k)} \\0 & 1 & -1 \\0 & 1 & 1 
\end{array}\right)^{T},
\end{aligned}
\end{equation}
respectively. 

In the full mixture model and GP-mixture model, we need to regress growth factors on the covariates with direct effects. We need to re-express the path coefficients if we reparameterize growth factors. The relationship between class-specific growth factor parameters in the original setting and those in the reparameterized frame can be further expressed with path coefficients as
\begin{equation}\nonumber
\setstretch{0.8}
\small
\begin{aligned}
&\boldsymbol{\mu}^{'(k)}_{\boldsymbol{\eta}}
=\boldsymbol{f}(\boldsymbol{\mu}^{(k)}_{\boldsymbol{\eta}})
\Longleftrightarrow E(\boldsymbol{\beta}^{'(k)}_{e0}+\boldsymbol{\beta}^{'(k)}_{e}\boldsymbol{x}_{ei}+\boldsymbol{\zeta}^{'}_{i})=\boldsymbol{f}(E(\boldsymbol{\beta}^{(k)}_{e0}+\boldsymbol{\beta}^{(k)}_{e}\boldsymbol{x}_{ei}+\boldsymbol{\zeta}_{i}))\\
&\Longleftrightarrow \boldsymbol{\beta}^{'(k)}_{e0}+\boldsymbol{\beta}^{'(k)}_{e}E(\boldsymbol{x}_{ei})=\boldsymbol{f}(\boldsymbol{\beta}^{(k)}_{e0}+\boldsymbol{\beta}^{(k)}_{e}E(\boldsymbol{x}_{ei}))\\
&\boldsymbol{\mu}^{(k)}_{\boldsymbol{\eta}}
=\boldsymbol{h}(\boldsymbol{\mu}^{'(k)}_{\boldsymbol{\eta}})
\Longleftrightarrow E(\boldsymbol{\beta}^{(k)}_{e0}+\boldsymbol{\beta}^{(k)}_{e}\boldsymbol{x}_{ei}+\boldsymbol{\zeta}_{i})=\boldsymbol{h}(E(\boldsymbol{\beta}^{'(k)}_{e0}+\boldsymbol{\beta}^{'(k)}_{e}\boldsymbol{x}_{ei}+\boldsymbol{\zeta}^{'}_{i}))\\
&\Longleftrightarrow \boldsymbol{\beta}^{(k)}_{e0}+\boldsymbol{\beta}^{(k)}_{e}E(\boldsymbol{x}_{ei})=\boldsymbol{h}(\boldsymbol{\beta}^{'(k)}_{e0}+\boldsymbol{\beta}^{'(k)}_{e}E(\boldsymbol{x}_{ei}))\\
&\boldsymbol{\Psi}^{'(k)}_{\boldsymbol{\eta}}=
\boldsymbol{\nabla_{f}}(\boldsymbol{\mu}^{(k)}_{\boldsymbol{\eta}})\boldsymbol{\Psi}^{(k)}_{\boldsymbol{\eta}}\boldsymbol{\nabla}^{T}_{\boldsymbol{f}}(\boldsymbol{\mu}^{(k)}_{\boldsymbol{\eta}})\\ &\Longleftrightarrow Var(\boldsymbol{\beta}^{'(k)}_{e0}+\boldsymbol{\beta}^{'(k)}_{e}\boldsymbol{x}_{ei}+\boldsymbol{\zeta}_{i}^{'})= \boldsymbol{\nabla_{f}}(\boldsymbol{\mu}^{(k)}_{\boldsymbol{\eta}})Var(\boldsymbol{\beta}^{(k)}_{e0}+\boldsymbol{\beta}^{(k)}_{e}\boldsymbol{x}_{ei}+\boldsymbol{\zeta}_{i})\boldsymbol{\nabla}^{T}_{\boldsymbol{f}}(\boldsymbol{\mu}^{(k)}_{\boldsymbol{\eta}})\\
&\Longleftrightarrow Var(\boldsymbol{\beta}^{'(k)}_{e}\boldsymbol{x}_{ei}+\boldsymbol{\zeta}_{i}^{'})=\boldsymbol{\nabla_{f}}(\boldsymbol{\mu}^{(k)}_{\boldsymbol{\eta}})Var(\boldsymbol{\beta}^{(k)}_{e}\boldsymbol{x}_{ei}+\boldsymbol{\zeta}_{i})\boldsymbol{\nabla}^{T}_{\boldsymbol{f}}(\boldsymbol{\mu}^{(k)}_{\boldsymbol{\eta}})\\
&\Longleftrightarrow \boldsymbol{\beta}^{'(k)}_{e}Var(\boldsymbol{x}_{ei})\boldsymbol{\beta}^{'(k)T}_{e}+Var(\boldsymbol{\zeta}_{i}^{'})=\boldsymbol{\nabla_{f}}(\boldsymbol{\mu}^{(k)}_{\boldsymbol{\eta}})\boldsymbol{\beta}^{(k)}_{e}Var(\boldsymbol{x}_{ei})\boldsymbol{\beta}^{(k)T}_{e}\boldsymbol{\nabla}^{T}_{\boldsymbol{f}}(\boldsymbol{\mu}^{(k)}_{\boldsymbol{\eta}})+\boldsymbol{\nabla_{f}}(\boldsymbol{\mu}^{(k)}_{\boldsymbol{\eta}})Var(\boldsymbol{\zeta}_{i})\boldsymbol{\nabla}^{T}_{\boldsymbol{f}}(\boldsymbol{\mu}^{(k)}_{\boldsymbol{\eta}})\ \ \ \ \ \ \ \ \ \ \ \ \ \\
&\Longleftrightarrow \boldsymbol{\beta}^{'(k)}_{e}=\boldsymbol{\nabla_{f}}(\boldsymbol{\mu}^{(k)}_{\boldsymbol{\eta}})\boldsymbol{\beta}^{(k)}_{e}\\
&\boldsymbol{\Psi}^{(k)}_{\boldsymbol{\eta}}=
\boldsymbol{\nabla_{h}}(\boldsymbol{\mu}^{'(k)}_{\boldsymbol{\eta}})\boldsymbol{\Psi}^{'(k)}_{\boldsymbol{\eta}}\boldsymbol{\nabla}^{T}_{\boldsymbol{h}}(\boldsymbol{\mu}^{'(k)}_{\boldsymbol{\eta}})\\ &\Longleftrightarrow Var(\boldsymbol{\beta}^{(k)}_{e0}+\boldsymbol{\beta}^{(k)}_{e}\boldsymbol{x}_{ei}+\boldsymbol{\zeta}_{i})=\boldsymbol{\nabla_{h}}(\boldsymbol{\mu}^{'(k)}_{\boldsymbol{\eta}})Var(\boldsymbol{\beta}^{'(k)}_{e0}+\boldsymbol{\beta}^{'(k)}_{e}\boldsymbol{x}_{ei}+\boldsymbol{\zeta}^{'}_{i})\boldsymbol{\nabla}^{T}_{\boldsymbol{h}}(\boldsymbol{\mu}^{'(k)}_{\boldsymbol{\eta}})\\
&\Longleftrightarrow Var(\boldsymbol{\beta}^{(k)}_{e}\boldsymbol{x}_{ei}+\boldsymbol{\zeta}_{i})=\boldsymbol{\nabla_{h}}(\boldsymbol{\mu}^{'(k)}_{\boldsymbol{\eta}})Var(\boldsymbol{\beta}^{'(k)}_{e}\boldsymbol{x}_{ei}+\boldsymbol{\zeta}^{'}_{i})\boldsymbol{\nabla}^{T}_{\boldsymbol{h}}(\boldsymbol{\mu}^{'(k)}_{\boldsymbol{\eta}})\nonumber\\
&\Longleftrightarrow \boldsymbol{\beta}^{(k)}_{e}Var(\boldsymbol{x}_{ei})\boldsymbol{\beta}^{(k)T}_{e}+Var(\boldsymbol{\zeta}_{i})=\boldsymbol{\nabla_{h}}(\boldsymbol{\mu}^{'(k)}_{\boldsymbol{\eta}})\boldsymbol{\beta}^{'(k)}_{e}Var(\boldsymbol{x}_{ei})\boldsymbol{\beta}^{'(k)T}_{e}\boldsymbol{\nabla}^{T}_{\boldsymbol{h}}(\boldsymbol{\mu}^{'(k)}_{\boldsymbol{\eta}})+\boldsymbol{\nabla_{h}}(\boldsymbol{\mu}^{'(k)}_{\boldsymbol{\eta}})Var(\boldsymbol{\zeta}^{'}_{i})\boldsymbol{\nabla}^{T}_{\boldsymbol{h}}(\boldsymbol{\mu}^{'(k)}_{\boldsymbol{\eta}})\\
&\Longleftrightarrow \boldsymbol{\beta}^{(k)}_{e}=\boldsymbol{\nabla_{h}}(\boldsymbol{\mu}^{'(k)}_{\boldsymbol{\eta}})\boldsymbol{\beta}^{'(k)}_{e}\nonumber
\end{aligned}
\end{equation}

\subsection{Model Specification and Estimation of GP-Mixture Models, CP-Mixture Models and Finite Mixture Models}\label{Supp:1C}
\subsubsection{Model Specification and Estimation of GP-Mixture Models}
The difference between GP-mixture models and full mixture models lies in that the membership of the GP-mixture model does not rely on any covariates. Accordingly, we need to modify Equation (\ref{eq:MoE_full}) which defines the full mixture model to be
\begin{equation}\label{eq:MoE_expert}
p(\boldsymbol{y}_{i}|z_{i}=k,\boldsymbol{x}_{ei})=\sum_{k=1}^{K}g(z_{i}=k)\times p(\boldsymbol{y}_{i}|z_{i}=k,\boldsymbol{x}_{ei}),
\end{equation}
where $g(z_{i}=k)$ can be viewed as the proportion of the sample in class $k$ with two constraints $0\le g(z_{i}=k)\le1$ and $\sum_{k=1}^{K}g(z_{i}=k)=1$. Equations (\ref{eq:MoE_expert}), (\ref{eq:Experts1}), and (\ref{eq:Experts2_full}) together define a GP-mixture model. 

From the definition, the within-class model of the GP-mixture model is the same as that of the full mixture model. Accordingly, the within-class implied mean vector ($\boldsymbol{\mu}_{i}^{(k)}$) and variance-covariance matrix ($\boldsymbol{\Sigma}_{i}^{(k)}$) of repeated outcomes $\boldsymbol{y}_{i}$ for the $i^{th}$ individual in the $k^{th}$ component are Equations (\ref{eq:Ymean_full}) and (\ref{eq:Yvar_full}), respectively. The parameters need to be estimated in the model specified in Equations (\ref{eq:MoE_expert}), (\ref{eq:Experts1}), and (\ref{eq:Experts2_full}) are given
\begin{equation}\nonumber
\begin{aligned}
\boldsymbol{\Theta}_{\text{GP}}&=\{\boldsymbol{\beta}_{e0}^{(k)}, \boldsymbol{\Psi}^{(k)}_{\boldsymbol{\eta}}, \gamma^{(k)}, \boldsymbol{\beta}_{e}^{(k)}, \boldsymbol{\mu}_{\boldsymbol{x}e}^{(k)}, \boldsymbol{\Phi}^{(k)}, \theta_{\epsilon}^{(k)}, \pi^{(k)}\}\\
&=\{\beta_{\eta_{0}}^{(k)}, \beta_{\eta_{1}}^{(k)}, \beta_{\eta_{2}}^{(k)}, \psi_{00}^{(k)}, \psi_{01}^{(k)}, \psi_{02}^{(k)}, \psi_{11}^{(k)}, \psi_{12}^{(k)}, \psi_{22}^{(k)}, \gamma^{(k)}, \boldsymbol{\beta}_{e}^{(k)}, \boldsymbol{\mu}_{\boldsymbol{x}e}^{(k)}, \boldsymbol{\Phi}^{(k)}, \theta_{\epsilon}^{(k)}, \pi^{(k)}\}\\
&k=2,\dots,K \text{ for }\pi^{(k)},\text{ indicating the proportion of the $k^{th}$ latent class},\\
&k=1,\dots,K \text{ for other parameters}.
\end{aligned}
\end{equation}
If we want to use the EM algorithm to obtain the estimates from the GP-mixture model, we also need to modify the cluster responsibilities at an iteration $t$ as
\begin{equation}\nonumber
\hat{r}_{ik}^{(t)}=\frac{\hat{\pi}^{(k)(t-1)}p(\boldsymbol{y}_{i}|\hat{\boldsymbol{\mu}}_{i}^{(k)(t-1)}, \hat{\boldsymbol{\Sigma}}_{i}^{(k)(t-1)}, \boldsymbol{x}_{ei})}{\sum_{k=1}^{K}\hat{\pi}^{(k)(t-1)}p(\boldsymbol{y}_{i}|\hat{\boldsymbol{\mu}}_{i}^{(k) (t-1)}, \hat{\boldsymbol{\Sigma}}_{i}^{(k) (t-1)}, \boldsymbol{x}_{ei})}.
\end{equation}

\subsubsection{Model Specification and Estimation of CP-Mixture Models}
The difference between CP-mixture models and full mixture models lies in that the class-specific growth factors do not depend on any covariates. Accordingly, we need to modify Equations (\ref{eq:MoE_full}) and (\ref{eq:Experts2_full}) that define a full mixture model to be
\begin{equation}\label{eq:MoE_gating}
p(\boldsymbol{y}_{i}|z_{i}=k,\boldsymbol{x}_{gi})=\sum_{k=1}^{K}g(z_{i}=k|\boldsymbol{x}_{gi})\times p(\boldsymbol{y}_{i}|z_{i}=k),
\end{equation}
and 
\begin{equation}\label{eq:Experts2_gating}
\boldsymbol{\eta}_{i}|(z_{i}=k)=\boldsymbol{\mu}^{(k)}_{\boldsymbol{\eta}}+\boldsymbol{\zeta}_{i},
\end{equation}
respectively. Equations (\ref{eq:MoE_gating}),  (\ref{eq:Gating_full}), (\ref{eq:Experts1}), and (\ref{eq:Experts2_gating}) together define a CP-mixture model. 

As the CP-mixture model's growth factors do not depend on any covariates, we need to remove the covariates from the model-implied mean vector and variance-covariance matrix and write them as
\begin{align}
&\boldsymbol{\mu}_{i}^{(k)}=\boldsymbol{\Lambda}_{i}(\gamma^{(k)})\boldsymbol{\mu}^{(k)}_{\boldsymbol{\eta}},\label{eq:Ymean_gating}\\
&\boldsymbol{\Sigma}_{i}^{(k)}=\boldsymbol{\Lambda}_{i}(\gamma^{(k)})\boldsymbol{\Psi}^{(k)}_{\boldsymbol{\eta}}\boldsymbol{\Lambda}_{i}(\gamma^{(k)})^{T}+\theta_{\epsilon}^{(k)}\boldsymbol{I}\label{eq:Yvar_gating},
\end{align}
where $\boldsymbol{\Psi}^{(k)}_{\boldsymbol{\eta}}$ is now defined as the variance-covariance matrix of the class-specific growth factors. The parameters need to be estimated in the model specified in Equations (\ref{eq:MoE_gating}), (\ref{eq:Gating_full}), (\ref{eq:Experts1}), and (\ref{eq:Experts2_gating}) are given
\begin{equation}\nonumber
\begin{aligned}
\boldsymbol{\Theta}_{\text{CP}}&=\{\boldsymbol{\mu}^{(k)}_{\boldsymbol{\eta}}, \boldsymbol{\Psi}^{(k)}_{\boldsymbol{\eta}}, \gamma^{(k)}, \theta_{\epsilon}^{(k)}, \beta_{g0}^{(k)}, \boldsymbol{\beta}_{g}^{(k)}\}\\
&=\{\mu_{\eta_{0}}^{(k)}, \mu_{\eta_{1}}^{(k)}, \mu_{\eta_{2}}^{(k)}, \psi_{00}^{(k)}, \psi_{01}^{(k)}, \psi_{02}^{(k)}, \psi_{11}^{(k)}, \psi_{12}^{(k)}, \psi_{22}^{(k)}, \gamma^{(k)}, \theta_{\epsilon}^{(k)}, \beta_{g0}^{(k)}, \boldsymbol{\beta}_{g}^{(k)}\},\\
&k=2,\dots,K \text{ for } \beta_{g0}^{(k)}, \boldsymbol{\beta}_{g}^{(k)},\\
&k=1,\dots,K \text{ for other parameters}.
\end{aligned}
\end{equation}
Cluster responsibilities at an iteration $t$ also need to be modified accordingly as
\begin{equation}\nonumber
\hat{r}_{ik}^{(t)}=\frac{g(z_{i}=k|\boldsymbol{x}_{gi})^{(t-1)}p(\boldsymbol{y}_{i}|\hat{\boldsymbol{\mu}}_{i}^{(k)(t-1)}, \hat{\boldsymbol{\Sigma}}_{i}^{(k)(t-1)})}{\sum_{k=1}^{K}g(z_{i}=k|\boldsymbol{x}_{gi})^{(t-1)}p(\boldsymbol{y}_{i}|\hat{\boldsymbol{\mu}}_{i}^{(k) (t-1)}, \hat{\boldsymbol{\Sigma}}_{i}^{(k) (t-1)})}.
\end{equation}

\subsubsection{Model Specification and Estimation of Finite Mixture Models}
The difference between finite mixture models and full mixture models lies in that neither the membership nor the class-specific growth factors depend on any covariates. Accordingly, we need to modify Equation (\ref{eq:MoE_full}) that defines a full mixture model to be 
\begin{equation}\label{eq:MoE_fmm}
p(\boldsymbol{y}_{i}|z_{i}=k)=\sum_{k=1}^{K}g(z_{i}=k)\times p(\boldsymbol{y}_{i}|z_{i}=k).
\end{equation}
Equations (\ref{eq:MoE_fmm}), (\ref{eq:Experts1}), and (\ref{eq:Experts2_gating}) together define a FMM. The within-class implied mean vector ($\boldsymbol{\mu}_{i}^{(k)}$) and variance-covariance matrix ($\boldsymbol{\Sigma}_{i}^{(k)}$) of a FMM can also be expressed as Equations (\ref{eq:Ymean_gating}) and (\ref{eq:Yvar_gating}), respectively. The parameters need to be estimated in the model specified in Equations (\ref{eq:MoE_fmm}), (\ref{eq:Experts1}), and (\ref{eq:Experts2_gating}) are given
\begin{equation}\nonumber
\begin{aligned}
\boldsymbol{\Theta}_{\text{FMM}}&=\{\boldsymbol{\mu}^{(k)}_{\boldsymbol{\eta}}, \boldsymbol{\Psi}^{(k)}_{\boldsymbol{\eta}}, \gamma^{(k)}, \theta_{\epsilon}^{(k)}, \pi^{(k)}\}\\
&=\{\mu_{\eta_{0}}^{(k)}, \mu_{\eta_{1}}^{(k)}, \mu_{\eta_{2}}^{(k)}, \psi_{00}^{(k)}, \psi_{01}^{(k)}, \psi_{02}^{(k)}, \psi_{11}^{(k)}, \psi_{12}^{(k)}, \psi_{22}^{(k)}, \gamma^{(k)}, \theta_{\epsilon}^{(k)}, \pi^{(k)}\},\\
&k=2,\dots,K \text{ for }\pi^{(k)},\text{ indicating the proportion of the $k^{th}$ latent class},\\
&k=1,\dots,K \text{ for other parameters}.
\end{aligned}
\end{equation}
Its cluster responsibilities at an iteration $t$ is
\begin{equation}\nonumber
\hat{r}_{ik}^{(t)}=\frac{\hat{\pi}^{(k)(t-1)}p(\boldsymbol{y}_{i}|\hat{\boldsymbol{\mu}}_{i}^{(k)(t-1)}, \hat{\boldsymbol{\Sigma}}_{i}^{(k)(t-1)})}{\sum_{k=1}^{K}\hat{\pi}^{(k)(t-1)}p(\boldsymbol{y}_{i}|\hat{\boldsymbol{\mu}}_{i}^{(k) (t-1)}, \hat{\boldsymbol{\Sigma}}_{i}^{(k) (t-1)})}.
\end{equation}

\newpage
\renewcommand\thetable{\arabic{table}}
\setcounter{table}{0}

\begin{table}
\centering
\begin{threeparttable}
\setlength{\tabcolsep}{5pt}
\renewcommand{\arraystretch}{0.75}
\caption{Performance Metrics: Definitions and Estimates}
\begin{tabular}{p{4cm}p{4.5cm}p{5.5cm}}
\hline
\hline
\textbf{Criteria} & \textbf{Definition} & \textbf{Estimate} \\
\hline
Relative Bias & $E_{\hat{\theta}}(\hat{\theta}-\theta)/\theta$ & $\sum_{s=1}^{S}(\hat{\theta}_{s}-\theta)/S\theta$ \\
Empirical SE & $\sqrt{Var(\hat{\theta})}$ & $\sqrt{\sum_{s=1}^{S}(\hat{\theta}_{s}-\bar{\theta})^{2}/(S-1)}$ \\
Relative RMSE & $\sqrt{E_{\hat{\theta}}(\hat{\theta}-\theta)^{2}}/\theta$ & $\sqrt{\sum_{s=1}^{S}(\hat{\theta}_{s}-\theta)^{2}/S}/\theta$ \\
Coverage Probability & $Pr(\hat{\theta}_{\text{low}}\le\theta\le\hat{\theta}_{\text{upper}})$ & $\sum_{s=1}^{S}I(\hat{\theta}_{\text{low},s}\le\theta\le\hat{\theta}_{\text{upper},s})/S$\\
\hline
\hline
\end{tabular}
\label{tbl:metric}
\begin{tablenotes}
\small
\item[1] {$\theta$: the population value of the parameter of interest} \\
\item[2] {$\hat{\theta}$: the estimate of $\theta$} \\
\item[3] {$S$: the number of replications and set as $1,000$ in our simulation study} \\
\item[4] {$s=1,\dots,S$: indexes the replications of the simulation} \\
\item[5] {$\hat{\theta}_{s}$: the estimate of $\theta$ from the $s^{th}$ replication} \\
\item[6] {$\bar{\theta}$: the mean of $\hat{\theta}_{s}$'s across replications} \\
\item[7] {$I()$: an indicator function}
\end{tablenotes}
\end{threeparttable}
\end{table}

\begin{table}
\centering
\begin{threeparttable}
\setlength{\tabcolsep}{5pt}
\renewcommand{\arraystretch}{0.75}
\caption{Simulation Design for the Mixture Model with the Bilinear Spline Growth Model with an Unknown Knot and Time-invariant Covariates as the Within-class Model in the Framework of Individual Measurement Occasions}
\begin{tabular}{p{3.9cm} p{12.5cm}}
\hline
\hline
\multicolumn{2}{c}{\textbf{Fixed Conditions}} \\
\hline
\textbf{Variables} & \textbf{Conditions} \\
\hline
Variance of Intercept & $\psi_{00}^{(k)}=25$ \\
\hline
Variance of Slopes & $\psi_{11}^{(k)}=\psi_{22}^{(k)}=1$ \\
\hline
Correlations of GFs\tnote{1} & $\rho^{(k)}=0.3$ \\
\hline
Time (\textit{t}) & $10$ scaled and equally spaced $t_{j} (j=0, \cdots, J-1, J=10)$ \\
\hline
Individual \textit{t} & $t_{ij} \sim U(t_{j}-\Delta, t_{j}+\Delta) (j=0, \cdots, J-1; \Delta=0.25)$ \\
\hline
Sample Size & $n=500$ \\
\hline
Mahalanobis distance & $d=0.86$ \\
\hline
\hline
\multicolumn{2}{c}{\textbf{Manipulated Conditions}} \\
\hline
\textbf{Variables} & \textbf{Conditions} \\
\hline
\multirow{3}{*}{Locations of knots} & $\mu_{\gamma}^{(1)}=4.00$; $\mu_{\gamma}^{(2)}=5.00$ \\
& $\mu_{\gamma}^{(1)}=3.75$; $\mu_{\gamma}^{(2)}=5.25$ \\
& $\mu_{\gamma}^{(1)}=3.50$; $\mu_{\gamma}^{(2)}=5.50$ \\
\hline
\multirow{2}{*}{Logistic Coefficients} & $\beta_{g0}=0$, $\beta_{g1}=\log(1.5)$, $\beta_{g2}=\log(1.7)$ (allocation ratio: $1$:$1$) \\
& $\beta_{g0}=0.775$, $\beta_{g1}=\log(1.5)$, $\beta_{g2}=\log(1.7)$ (allocation ratio: $1$:$2$) \\
\hline
\multirow{3}{*}{Path Coefficients\tnote{2}} & Covariates explain $13\%$ variability of GFs in both clusters \\
& Covariates explain $13\%$ and $26\%$ variability of GFs in Cluster $1$ and $2$ \\
& Covariates explain $26\%$ variability of GFs in both clusters \\
\hline
Residual Variance & $\theta_{\epsilon}^{(k)}=1 \text{ or } 2$ \\
\hline
\hline
\multicolumn{2}{l}{\textbf{Scenario 1: Different means of initial status and (means of) knot locations}} \\
\hline
\textbf{Variables} & \textbf{Conditions of 2 latent classes} \\
\hline
Means of Slope 1's & $\mu_{\eta_{1}}^{(k)}=-5$ $(k=1, 2)$ \\
\hline
Means of Slope 2's
& $\mu_{\eta_{2}}^{(k)}=-2.6$ $(k=1, 2)$ \\
\hline
Means of Intercepts & $\mu_{\eta_{0}}^{(1)}=98$, $\mu_{\eta_{0}}^{(2)}=102$ $(d=0.86)$ \\
\hline
\hline
\multicolumn{2}{l}{\textbf{Scenario 2: Different means of slope 1 and (means of) knot locations}} \\
\hline
\textbf{Variables} & \textbf{Conditions of 2 latent classes} \\
\hline
Means of Intercepts & $\mu_{\eta_{0}}^{(k)}=100$ $(k=1,2)$ \\
\hline
Means of Slope 2's & $\mu_{\eta_{2}}^{(k)}=-2$ $(k=1, 2)$ \\
\hline
Means of Slope 1's & $\mu_{\eta_{1}}^{(1)}=-4.4$, $\mu_{\eta_{1}}^{(2)}=-3.6$ $(d=0.86)$ \\
\hline
\hline
\multicolumn{2}{l}{\textbf{Scenario 3: Different means of slope 2 and (means of) knot locations}} \\
\hline
\textbf{Variables} & \textbf{Conditions of 2 latent classes} \\
\hline
Means of Intercepts & $\mu_{\eta_{0}}^{(k)}=100$ $(k=1,2)$ \\
\hline
Means of Slope 1's & $\mu_{\eta_{1}}^{(k)}=-5$ $(k=1,2)$ \\
\hline
Means of Slope 2's & $\mu_{\eta_{2}}^{(1)}=-2.6$, $\mu_{\eta_{2}}^{(2)}=-3.4$ $(d=0.86)$ \\
\hline
\hline
\end{tabular}
\label{tbl:simu}
\begin{tablenotes}
\small
\item[1] {GFs represent Growth Factors.}\\
\item[2] {For each class-specific path coefficients, $\beta_{e2}=1.5\beta_{e1}$.}
\end{tablenotes}
\end{threeparttable}
\end{table}

\begin{table}
\centering
\resizebox{1.15\textwidth}{!}{
\begin{threeparttable}
\setlength{\tabcolsep}{5pt}
\renewcommand{\arraystretch}{0.75}
\caption{Scenarios of Demonstrating the Approach for Identifying Covariates}
\begin{tabular}{p{1cm} C{6.5cm}C{6.5cm}C{5cm}}
\hline
\hline
Scenario & Latent Class $1$ & Latent Class $2$ & Multinomial Function \\
\hline
$1$ & Covariates explain $\ \ 2\%$ variability\tnote{1} & ---\tnote{2} & --- \\
$2$ & Covariates explain $13\%$ variability & --- & --- \\
$3$ & Covariates explain $26\%$ variability & --- & --- \\
$4$ & Covariates explain $\ \ 2\%$ variability & Covariates explain $\ \ 2\%$ variability & $\beta_{1}=\log(1.5)$, $\beta_{2}=\log(1.7)$ \\
$5$ & Covariates explain $\ \ 2\%$ variability & Covariates explain $13\%$ variability & $\beta_{1}=\log(1.5)$, $\beta_{2}=\log(1.7)$ \\
$6$ & Covariates explain $13\%$ variability & Covariates explain $13\%$ variability & $\beta_{1}=\log(1.5)$, $\beta_{2}=\log(1.7)$ \\
$7$ & Covariates explain $13\%$ variability & Covariates explain $26\%$ variability & $\beta_{1}=\log(1.5)$, $\beta_{2}=\log(1.7)$ \\
$8$ & Covariates explain $26\%$ variability & Covariates explain $26\%$ variability & $\beta_{1}=\log(1.5)$, $\beta_{2}=\log(1.7)$ \\
\hline
\hline
\end{tabular}
\label{tbl:approach}
\begin{tablenotes}
    \small
\item[1] For the path coefficients in each latent class, we set $\beta_{e2}=1.5\beta_{e1}$.\\
\item[2] --- indicates that the corresponding metric is not applicable for that scenario.\\
\end{tablenotes}
\end{threeparttable}}
\end{table}

\begin{table}
\centering
\begin{threeparttable}
\setlength{\tabcolsep}{5pt}
\renewcommand{\arraystretch}{0.75}
\caption{Mixture Models Constructed for Analyzing the Motivating Data}
\begin{tabular}{lrrrrrrr}
\hline
\hline
Model\tnote{1} & -2ll & AIC & BIC & Entropy & Kappa Statistic\tnote{2} & Judgment\tnote{3} & Time\tnote{4} \\
\hline
Model $1$ & $31077$ & $31147$ & $31294$ & $0.654$ &  ---\tnote{5} & --- & $7.75$ min\\
Model $2$ & $31009$ & $31095$ & $31276$ & $0.684$ & $0.83$ ($0.79$, $0.88$) & Almost perfect agreement & $11.65$ min \\
Model $3$ & $30990$ & $31084$ & $31282$ & $0.694$ & $0.84$ ($0.80$, $0.88$) & Almost perfect agreement & $15.44$ min\\
Model $4$ & $33342$ & $33478$ & $33765$ & $0.682$ & $0.85$ ($0.81$, $0.89$) & Almost perfect agreement & $38.38$ min\\
Model $5$ & $33204$ & $33340$ & $33627$ & $0.788$ & $0.11$ ($0.05$, $0.17$) & Slight agreement & $41.77$ min \\
Model $6$ & $33282$ & $33434$ & $33755$ & $0.699$ & $0.87$ ($0.83$, $0.91$) & Almost perfect agreement & $39.12$ min\\
\hline
\hline
\end{tabular}
\label{tbl:compare}
\begin{tablenotes}
\small
\item[1] Model $1$---Model $6$ are (1) a finite mixture model without any covariates, (2) a CP-mixture model with family income, parents' highest education, sex and race/ethnicity in the multinomial functions, (3) a CP-mixture model with family income, parents' highest education, sex, race/ethnicity,  attentional-focus, approach-to-learning in all multinomial functions, (4) a GP-mixture model with the variable attentional-focus and approach-to-learning in all within-class models, (5) a GP-mixture model with the variable family income and parents' highest education in all within-class models and (6) a full mixture with the covariate attentional-focus in all within-class models and the other four covariates in all multinomial functions. All models are constructed with $3$ latent classes.\\
\item[2] Kappa statistic, which is Dumenci's Latent Kappa coefficient, is for the agreement between the membership obtained from the FMM and each of the other models.\\
\item[3] Judgment is based on the output of the \textit{R} package \textit{fmsb}.\\
\item[4] Time taken is the CPU time charged for the execution of each model, which is recorded by the \textit{R} function \textit{proc.time()}.\\
\item[5] --- indicates that the corresponding metric is not applicable for the FMM.
\end{tablenotes}
\end{threeparttable}
\end{table}

\begin{table}
\centering
\resizebox{1.15\textwidth}{!}{
\begin{threeparttable}
\setlength{\tabcolsep}{5pt}
\renewcommand{\arraystretch}{0.75}
\caption{Estimates of Growth Mixture Model with Bilinear Spline Change Patterns (3 Latent Classes)}
\begin{tabular}{lrrrrrr}
\hline
\hline
& \multicolumn{2}{c}{\textbf{Class 1}} & \multicolumn{2}{c}{\textbf{Class 2}}& \multicolumn{2}{c}{\textbf{Class 3}} \\
\hline
\textbf{Mean of Growth Factor} & Estimate (SE) & P value & Estimate (SE) & P value & Estimate (SE) & P value \\
\hline
\textbf{Intercept}\tnote{1} & $24.046$ ($1.816$) & $<0.0001^{\ast}$\tnote{2} & $22.037$ ($0.893$) & $<0.0001^{\ast}$ & $32.092$ ($1.013$) & $<0.0001^{\ast}$ \\
\textbf{Slope $1$} & $1.921$ ($0.087$) & $<0.0001^{\ast}$ & $1.684$ ($0.028$) & $<0.0001^{\ast}$ & $2.075$ ($0.031$) & $<0.0001^{\ast}$ \\
\textbf{Slope $2$} & $0.925$ ($0.033$) & $<0.0001^{\ast}$ & $0.554$ ($0.037$) & $<0.0001^{\ast}$ & $0.675$ ($0.025$) & $<0.0001^{\ast}$ \\
\hline
\hline
\textbf{Additional Parameter} & Estimate (SE) & P value & Estimate (SE) & P value & Estimate (SE) & P value \\
\hline
\textbf{Knot} & $82.238$ ($0.811$) & $<0.0001^{\ast}$ & $108.929$ ($0.659$) & $<0.0001^{\ast}$ & $97.567$ ($0.539$) & $<0.0001^{\ast}$ \\
\hline
\hline
\textbf{Variance of Growth Factor} & Estimate (SE) & P value & Estimate (SE) & P value & Estimate (SE) & P value \\
\hline
\textbf{Intercept} & $94.919$ ($27.720$) & $0.0006^{\ast}$ & $65.310$ ($11.906$) & $<0.0001^{\ast}$ & $85.261$ ($14.567$) & $<0.0001^{\ast}$ \\
\textbf{Slope $1$} & $0.241$ ($0.078$) & $0.002^{\ast}$ & $0.051$ ($0.011$) & $<0.0001^{\ast}$ & $0.037$ ($0.010$) & $0.0002^{\ast}$ \\
\textbf{Slope $2$} & $0.038$ ($0.009$) & $<0.0001^{\ast}$ & $0.028$ ($0.013$) & $0.0313^{\ast}$ & $0.008$ ($0.006$) & $0.1824$ \\
\hline
\hline
\end{tabular}
\label{tbl:FMM_est}
\begin{tablenotes}
\small
\item[1] Intercept was defined as mathematics IRT scores at 60-month old in this case.\\
\item[2] $^{\ast}$ indicates statistical significance at $0.05$ level.\\
\end{tablenotes}
\end{threeparttable}}
\end{table}

\begin{table}
\centering
\resizebox{1.15\textwidth}{!}{
\begin{threeparttable}
\setlength{\tabcolsep}{5pt}
\renewcommand{\arraystretch}{0.75}
\caption{Estimates of CP-mixture model with Bilinear Spline Change Patterns (3 Latent Classes)}
\begin{tabular}{lrrrrrr}
\hline
\hline
& \multicolumn{2}{c}{\textbf{Class 1}} & \multicolumn{2}{c}{\textbf{Class 2}}& \multicolumn{2}{c}{\textbf{Class 3}} \\
\hline
\textbf{Mean of Growth Factor} & Estimate (SE) & P value & Estimate (SE) & P value & Estimate (SE) & P value \\
\hline
\textbf{Intercept}\tnote{1} & $22.498$ ($1.930$) & $<0.0001^{\ast}$\tnote{2} & $21.706$ ($0.876$) & $<0.0001^{\ast}$ & $32.176$ ($1.101$) & $<0.0001^{\ast}$ \\
\textbf{Slope $1$} & $1.872$ ($0.078$) & $<0.0001^{\ast}$ & $1.667$ ($0.029$) & $<0.0001^{\ast}$ & $2.066$ ($0.028$) & $<0.0001^{\ast}$ \\
\textbf{Slope $2$} & $0.940$ ($0.029$) & $<0.0001^{\ast}$ & $0.508$ ($0.039$) & $<0.0001^{\ast}$ & $0.680$ ($0.023$) & $<0.0001^{\ast}$ \\
\hline
\hline
\textbf{Additional Parameter} & Estimate (SE) & P value & Estimate (SE) & P value & Estimate (SE) & P value \\
\hline
\textbf{Knot} & $82.190$ ($0.022$) & $<0.0001^{\ast}$ & $110.029$ ($0.682$) & $<0.0001^{\ast}$ & $97.868$ ($0.553$) & $<0.0001^{\ast}$ \\
\hline
\hline
\textbf{Variance of Growth Factor} & Estimate (SE) & P value & Estimate (SE) & P value & Estimate (SE) & P value \\
\hline
\textbf{Intercept} & $76.159$ ($28.955$) & $0.0085^{\ast}$ & $60.347$ ($10.773$) & $<0.0001^{\ast}$ & $100.901$ ($16.245$) & $<0.0001^{\ast}$ \\
\textbf{Slope $1$} & $0.182$ ($0.066$) & $0.0058^{\ast}$ & $0.042$ ($0.009$) & $<0.0001^{\ast}$ & $0.039$ ($0.010$) & $0.0001^{\ast}$ \\
\textbf{Slope $2$} & $0.037$ ($0.008$) & $<0.0001^{\ast}$ & $0.033$ ($0.015$) & $0.0278^{\ast}$ & $0.010$ ($0.006$) & $0.0956$ \\
\hline
\hline
\textbf{Logistic Coef.} & \multicolumn{2}{r}{OR ($95\%$ CI)} & \multicolumn{2}{r}{OR ($95\%$ CI)} & \multicolumn{2}{r}{OR ($95\%$ CI)} \\
\hline
\textbf{Family Income} & \multicolumn{2}{r}{---\tnote{3}} & \multicolumn{2}{r}{$0.970$ ($0.894$, $1.052$)} & \multicolumn{2}{r}{$1.030$ ($0.947$, $1.121$)} \\
\textbf{Parents' Highest Education} & \multicolumn{2}{r}{---} & \multicolumn{2}{r}{$0.979$ ($0.774$, $1.238$)} & \multicolumn{2}{r}{$1.386$ ($1.095$, $1.753$)$^{\ast}$} \\
\textbf{Sex}$ ($0---Boy; 1---Girl) & \multicolumn{2}{r}{---} & \multicolumn{2}{r}{$0.391$ ($0.187$, $0.816$)$^{\ast}$} & \multicolumn{2}{r}{$0.379$ ($0.185$, $0.776$)$^{\ast}$} \\
\textbf{Race}$ ($0---White; 1---Others) & \multicolumn{2}{r}{---} & \multicolumn{2}{r}{$0.586$ ($0.279$, $1.232$)} & \multicolumn{2}{r}{$0.446$ ($0.220$, $0.904$)$^{\ast}$} \\
\hline
\hline
\end{tabular}
\label{tbl:Gating_est}
\begin{tablenotes}
\small
\item[1] Intercept was defined as mathematics IRT scores at 60-month old in this case.\\
\item[2] $^{\ast}$ indicates statistical significance at $0.05$ level.\\
\item[3] We set Class $1$ as the reference group.\\
\end{tablenotes}
\end{threeparttable}}
\end{table}

\begin{table}
\centering
\resizebox{1.15\textwidth}{!}{
\begin{threeparttable}
\setlength{\tabcolsep}{5pt}
\renewcommand{\arraystretch}{0.75}
\caption{Estimates of GP-mixture model with Bilinear Spline Change Patterns (3 Latent Classes)}
\begin{tabular}{lrrrrrr}
\hline
\hline
& \multicolumn{2}{c}{\textbf{Class 1}} & \multicolumn{2}{c}{\textbf{Class 2}}& \multicolumn{2}{c}{\textbf{Class 3}} \\
\hline
\textbf{Intercept of Growth Factor} & Estimate (SE) & P value & Estimate (SE) & P value & Estimate (SE) & P value \\
\hline
\textbf{Intercept}\tnote{1} & $24.069$ ($1.841$) & $<0.0001^{\ast}$\tnote{2} & $23.008$ ($0.729$) & $<0.0001^{\ast}$ & $32.067$ ($1.064$) & $<0.0001^{\ast}$ \\
\textbf{Slope $1$} & $2.062$ ($0.092$) & $<0.0001^{\ast}$ & $1.651$ ($0.027$) & $<0.0001^{\ast}$ & $2.080$ ($0.031$) & $<0.0001^{\ast}$ \\
\textbf{Slope $2$} & $0.858$ ($0.037$) & $<0.0001^{\ast}$ & $0.590$ ($0.032$) & $<0.0001^{\ast}$ & $0.665$ ($0.026$) & $<0.0001^{\ast}$ \\
\hline
\hline
\textbf{Additional Parameter} & Estimate (SE) & P value & Estimate (SE) & P value & Estimate (SE) & P value \\
\hline
\textbf{Knot} & $83.670$ ($0.034$) & $<0.0001^{\ast}$ & $108.648$ ($0.647$) & $<0.0001^{\ast}$ & $97.555$ ($0.550$) & $<0.0001^{\ast}$ \\
\hline
\hline
\textbf{Residual of Growth Factor} & Estimate (SE) & P value & Estimate (SE) & P value & Estimate (SE) & P value \\
\hline
\textbf{Intercept} & $66.657$ ($22.252$) & $0.0027^{\ast}$ & $62.036$ ($9.303$) & $<0.0001^{\ast}$ & $88.825$ ($13.816$) & $<0.0001^{\ast}$ \\
\textbf{Slope $1$} & $0.142$ ($0.054$) & $0.0085^{\ast}$ & $0.070$ ($0.010$) & $<0.0001^{\ast}$ & $0.038$ ($0.010$) & $0.0001^{\ast}$ \\
\textbf{Slope $2$} & $0.024$ ($0.008$) & $0.0027^{\ast}$ & $0.039$ ($0.012$) & $0.0012^{\ast}$ & $0.005$ ($0.006$) & $0.4047$ \\
\hline
\hline
\textbf{Mean of Covariates}\tnote{3} & Estimate (SE) & P value & Estimate (SE) & P value & Estimate (SE) & P value \\
\hline
\textbf{attentional-focus} & $-0.513$ ($0.183$) & $0.0051^{\ast}$ & $-0.025$ ($0.079$) & $0.7517$ & $0.240$ ($0.083$) & $0.0038^{\ast}$ \\
\textbf{approach-to-learning} & $-0.446$ ($0.169$) & $0.0083^{\ast}$ & $-0.052$ ($0.081$) & $0.5209$ & $0.249$ ($0.079$) & $0.0016^{\ast}$ \\
\hline
\hline
\textbf{Variances of Covariates} & Estimate (SE) & P value & Estimate (SE) & P value & Estimate (SE) & P value \\
\hline
\textbf{attentional-focus} & $1.407$ ($0.266$) & $<0.0001^{\ast}$ & $0.920$ ($0.095$) & $<0.0001^{\ast}$ & $0.770$ ($0.100$) & $<0.0001^{\ast}$ \\
\textbf{approach-to-learning} & $1.061$ ($0.211$) & $<0.0001^{\ast}$ & $1.038$ ($0.102$) & $<0.0001^{\ast}$ & $0.775$ ($0.090$) & $<0.0001^{\ast}$ \\
\hline
\hline
\textbf{Path Coefficients} & Estimate (SE) & P value & Estimate (SE) & P value & Estimate (SE) & P value \\
\hline
\textbf{attentional-focus to Intercept} & $1.032$ ($1.959$) & $0.5983$ & $1.605$ ($1.052$) & $0.1271$ & $1.887$ ($1.637$) & $0.2490$ \\
\textbf{attentional-focus to Slope $1$} & $0.221$ ($0.087$) & $0.0111^{\ast}$ & $0.064$ ($0.034$) & $0.0598$ & $-0.010$ ($0.043$) & $0.8161$ \\
\textbf{attentional-focus  to Slope $2$} & $-0.171$ ($0.046$) & $0.0002^{\ast}$ & $0.056$ ($0.038$) & $0.1406$ & $-0.030$ ($0.034$) & $0.3776$ \\
\textbf{approach-to-learning to Intercept} & $1.255$ ($2.180$) & $0.5648$ & $2.337$ ($0.964$) & $0.0153$ & $-0.267$ ($1.600$) & $0.8675$ \\
\textbf{approach-to-learning to Slope $1$} & $-0.058$ ($0.092$) & $0.5284$ & $-0.058$ ($0.031$) & $0.0613$ & $-0.042$ ($0.042$) & $0.3173$ \\
\textbf{approach-to-learning to Slope $2$} & $0.119$ ($0.053$) & $0.0247^{\ast}$ & $-0.054$ ($0.035$) & $0.1229$ & $0.067$ ($0.032$) & $0.0363^{\ast}$ \\
\hline
\hline
\end{tabular}
\label{tbl:Expert_est}
\begin{tablenotes}
\small
\item[1] Intercept was defined as mathematics IRT scores at 60-month old in this case.\\
\item[2] $^{\ast}$ indicates statistical significance at $0.05$ level.\\
\item[3] We built the model using standardized covariate with direct effects (i.e., attentional-focus and approach-to-learning).\\
\end{tablenotes}
\end{threeparttable}}
\end{table}

\begin{table}
\centering
\resizebox{1.15\textwidth}{!}{
\begin{threeparttable}
\setlength{\tabcolsep}{5pt}
\renewcommand{\arraystretch}{0.75}
\caption{Estimates of Full Mixture Model with Bilinear Spline Change Patterns (3 Latent Classes)}
\begin{tabular}{lrrrrrr}
\hline
\hline
& \multicolumn{2}{c}{\textbf{Class 1}} & \multicolumn{2}{c}{\textbf{Class 2}}& \multicolumn{2}{c}{\textbf{Class 3}} \\
\hline
\textbf{Intercept of Growth Factor} & Estimate (SE) & P value & Estimate (SE) & P value & Estimate (SE) & P value \\
\hline
\textbf{Intercept}\tnote{1} & $20.896$ ($1.455$) & $<0.0001^{\ast}$\tnote{2} & $22.752$ ($0.800$) & $<0.0001^{\ast}$ & $32.765$ ($0.987$) & $<0.0001^{\ast}$ \\
\textbf{Slope $1$} & $1.947$ ($0.068$) & $<0.0001^{\ast}$ & $1.671$ ($0.026$) & $<0.0001^{\ast}$ & $2.077$ ($0.030$) & $<0.0001^{\ast}$ \\
\textbf{Slope $2$} & $0.899$ ($0.040$) & $<0.0001^{\ast}$ & $0.577$ ($0.035$) & $<0.0001^{\ast}$ & $0.682$ ($0.023$) & $<0.0001^{\ast}$ \\
\hline
\hline
\textbf{Additional Parameter} & Estimate (SE) & P value & Estimate (SE) & P value & Estimate (SE) & P value \\
\hline
\textbf{Knot} & $82.190$ ($0.021$) & $<0.0001^{\ast}$ & $108.753$ ($0.635$) & $<0.0001^{\ast}$ & $97.139$ ($0.487$) & $<0.0001^{\ast}$ \\
\hline
\hline
\textbf{Residual of Growth Factor} & Estimate (SE) & P value & Estimate (SE) & P value & Estimate (SE) & P value \\
\hline
\textbf{Intercept} & $28.890$ ($13.486$) & $0.0322^{\ast}$ & $62.819$ ($10.103$) & $<0.0001^{\ast}$ & $97.677$ ($14.122$) & $<0.0001^{\ast}$ \\
\textbf{Slope $1$} & $0.054$ ($0.041$) & $0.1878$ & $0.059$ ($0.010$) & $<0.0001^{\ast}$ & $0.041$ ($0.010$) & $<0.0001^{\ast}$ \\
\textbf{Slope $2$} & $0.037$ ($0.010$) & $0.0002^{\ast}$ & $0.041$ ($0.013$) & $0.0016^{\ast}$ & $0.006$ ($0.006$) & $0.3173$ \\
\hline
\hline
\textbf{Mean of Covariates}\tnote{3} & Estimate (SE) & P value & Estimate (SE) & P value & Estimate (SE) & P value \\
\hline
\textbf{attentional-focus} & $-0.325$ ($0.201$) & $0.1059$ & $-0.074$ ($0.090$) & $0.4110$ & $0.207$ ($0.079$) & $0.0088^{\ast}$ \\
\textbf{approach-to-learning} & $-0.232$ ($0.185$) & $0.2098$ & $-0.108$ ($0.090$) & $0.2301$ & $0.212$ ($0.074$) & $0.0042^{\ast}$ \\
\hline
\hline
\textbf{Variances of Covariates} & Estimate (SE) & P value & Estimate (SE) & P value & Estimate (SE) & P value \\
\hline
\textbf{attentional-focus} & $1.297$ ($0.246$) & $<0.0001^{\ast}$ & $0.969$ ($0.110$) & $<0.0001^{\ast}$ & $0.832$ ($0.101$) & $<0.0001^{\ast}$ \\
\textbf{approach-to-learning} & $1.134$ ($0.207$) & $<0.0001^{\ast}$ & $1.063$ ($0.108$) & $<0.0001^{\ast}$ & $0.794$ ($0.087$) & $<0.0001^{\ast}$ \\
\hline
\hline
\textbf{Path Coefficients} & Estimate (SE) & P value & Estimate (SE) & P value & Estimate (SE) & P value \\
\hline
\textbf{attentional-focus to Intercept} & $-0.873$ ($1.656$) & $0.5981$ & $1.828$ ($1.174$) & $0.1195$ & $1.253$ ($1.733$) & $0.4697$ \\
\textbf{attentional-focus to Slope $1$} & $0.333$ ($0.096$) & $0.0005^{\ast}$ & $0.028$ ($0.032$) & $0.3816$ & $0.019$ ($0.042$) & $0.6510$ \\
\textbf{attentional-focus  to Slope $2$} & $-0.058$ ($0.049$) & $0.2365$ & $0.046$ ($0.041$) & $0.2619$ & $-0.047$ ($0.035$) & $0.1793$ \\
\textbf{approach-to-learning to Intercept} & $3.641$ ($1.829$) & $0.0465^{\ast}$ & $1.546$ ($1.043$) & $0.1383$ & $-0.094$ ($1.743$) & $0.9570$ \\
\textbf{approach-to-learning to Slope $1$} & $-0.182$ ($0.098$) & $0.0633$ & $-0.014$ ($0.029$) & $0.6293$ & $-0.054$ ($0.043$) & $0.2092$ \\
\textbf{approach-to-learning to Slope $2$} & $0.021$ ($0.048$) & $0.6617$ & $-0.052$ ($0.037$) & $0.1599$ & $0.072$ ($0.036$) & $0.0455^{\ast}$ \\
\hline
\hline
\textbf{Logistic Coef.} & \multicolumn{2}{r}{OR ($95\%$ CI)} & \multicolumn{2}{r}{OR ($95\%$ CI)} & \multicolumn{2}{r}{OR ($95\%$ CI)} \\
\hline
\textbf{Family Income} & \multicolumn{2}{r}{---\tnote{4}} & \multicolumn{2}{r}{$0.993$ ($0.909$, $1.084$)} & \multicolumn{2}{r}{$1.046$ ($0.957$, $1.144$)} \\
\textbf{Parents' Highest Education} & \multicolumn{2}{r}{---} & \multicolumn{2}{r}{$1.015$ ($0.800$, $1.287$)} & \multicolumn{2}{r}{$1.391$ ($1.090$, $1.774$)$^{\ast}$} \\
\textbf{Sex}$ ($0---Boy; 1---Girl) & \multicolumn{2}{r}{---} & \multicolumn{2}{r}{$0.449$ ($0.196$, $1.027$)} & \multicolumn{2}{r}{$0.412$ ($0.195$, $0.872$)$^{\ast}$} \\
\textbf{Race}$ ($0---White; 1---Others) & \multicolumn{2}{r}{---} & \multicolumn{2}{r}{$0.610$ ($0.291$, $1.281$)} & \multicolumn{2}{r}{$0.447$ ($0.218$, $0.917$)$^{\ast}$} \\
\hline
\hline
\end{tabular}
\label{tbl:full_est}
\begin{tablenotes}
\small
\item[1] Intercept was defined as mathematics IRT scores at 60-month old in this case.\\
\item[2] $^{\ast}$ indicates statistical significance at $0.05$ level.\\
\item[3] We built the model using standardized covariate with direct effects (i.e., attentional-focus and approach-to-learning).\\
\item[4] We set Class $1$ as the reference group.\\
\end{tablenotes}
\end{threeparttable}}
\end{table}

\renewcommand\thefigure{\arabic{figure}}
\setcounter{figure}{0}
\begin{figure}
\centering
\includegraphics[width=0.8\textwidth]{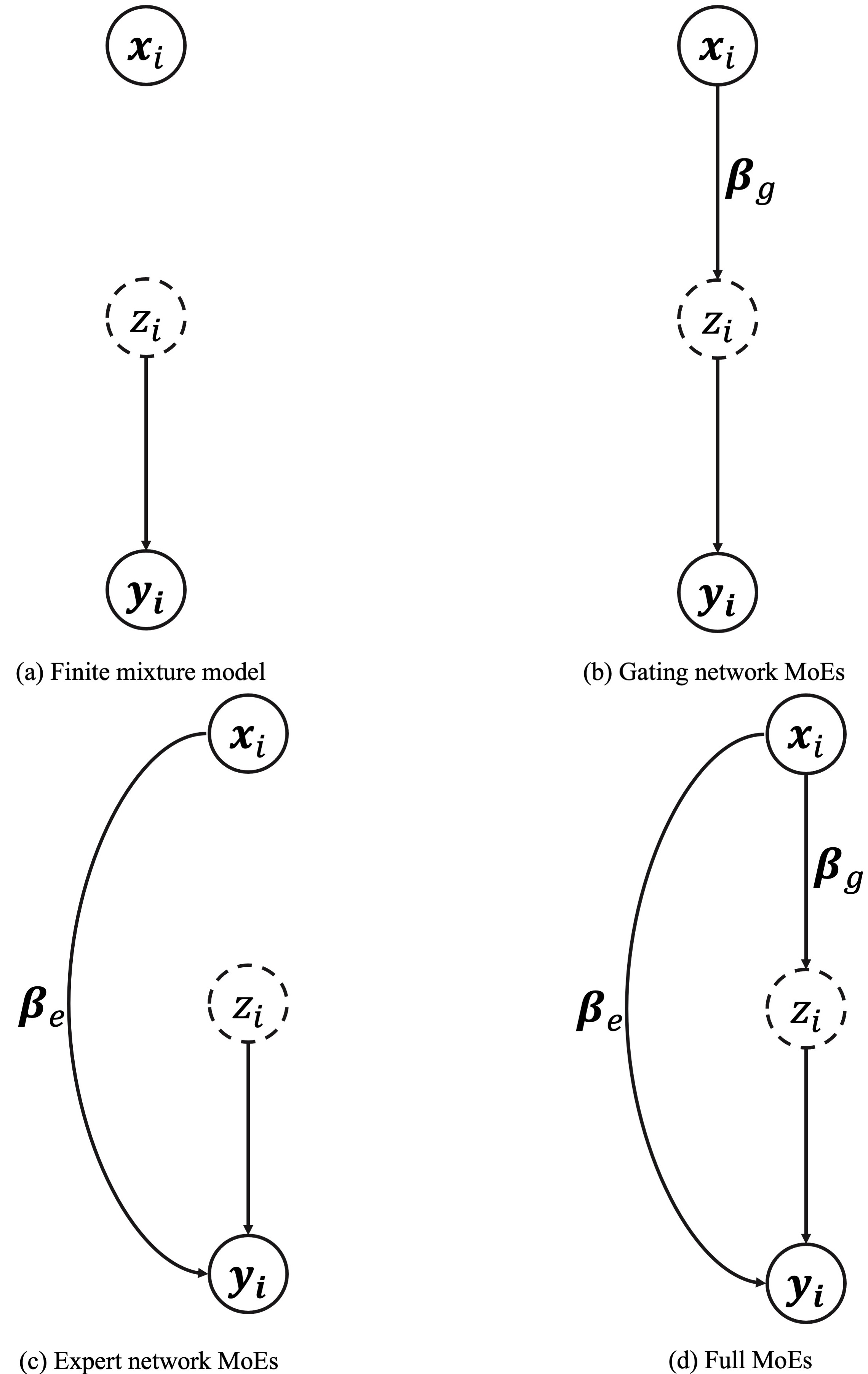}
\caption{The Graphic Model Representation of the Full and Reduced Mixture-of-Experts Models}
\label{fig:MoE}
\end{figure}

\begin{figure}
\centering
\includegraphics[width=0.98\textwidth]{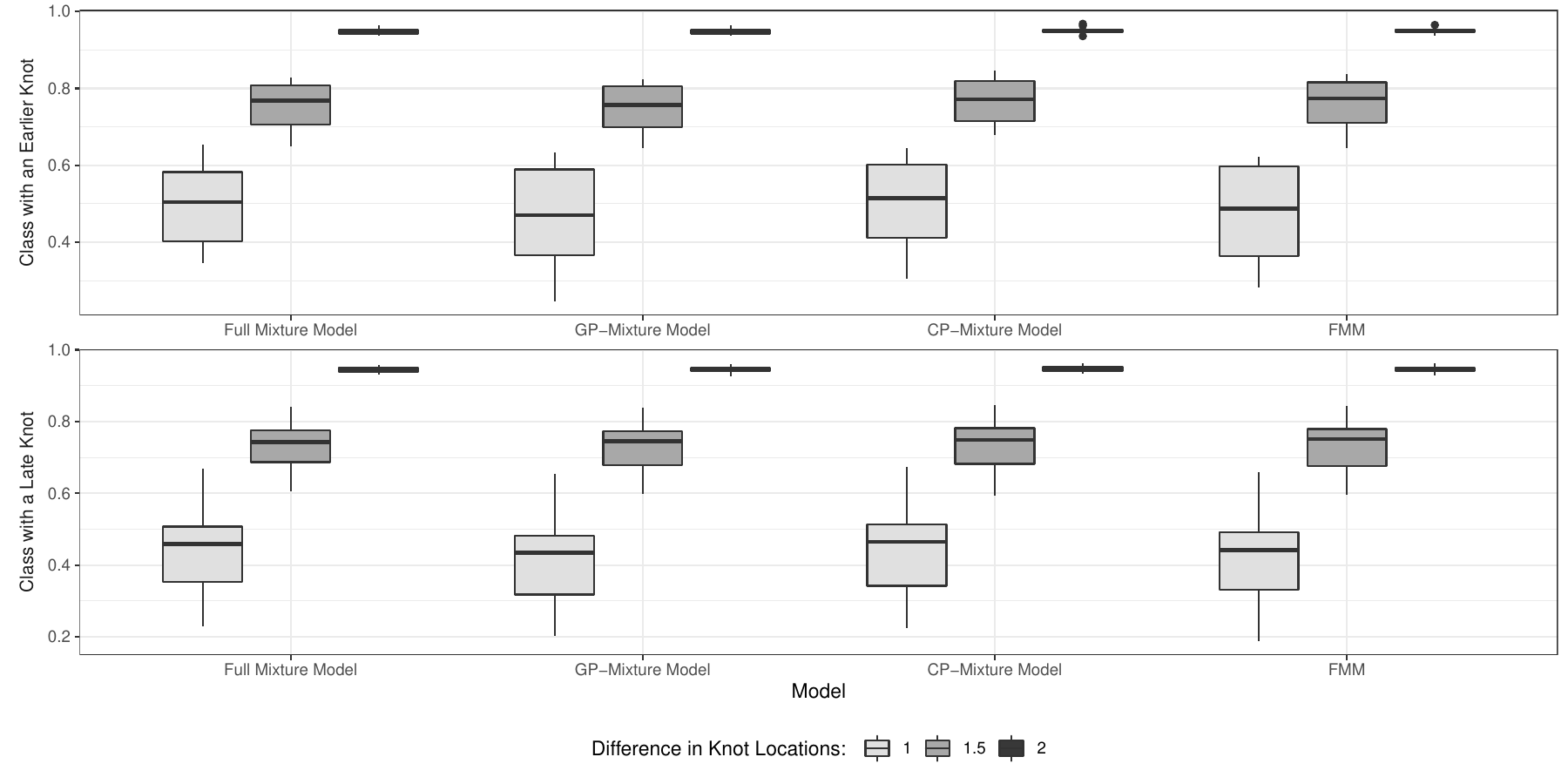}
\caption{Coverage Probabilities of Class-Specific Knot of Proposed Mixture Models}
\label{fig:KnotCP}
\end{figure}

\begin{figure}
\centering
\includegraphics[width=0.98\textwidth]{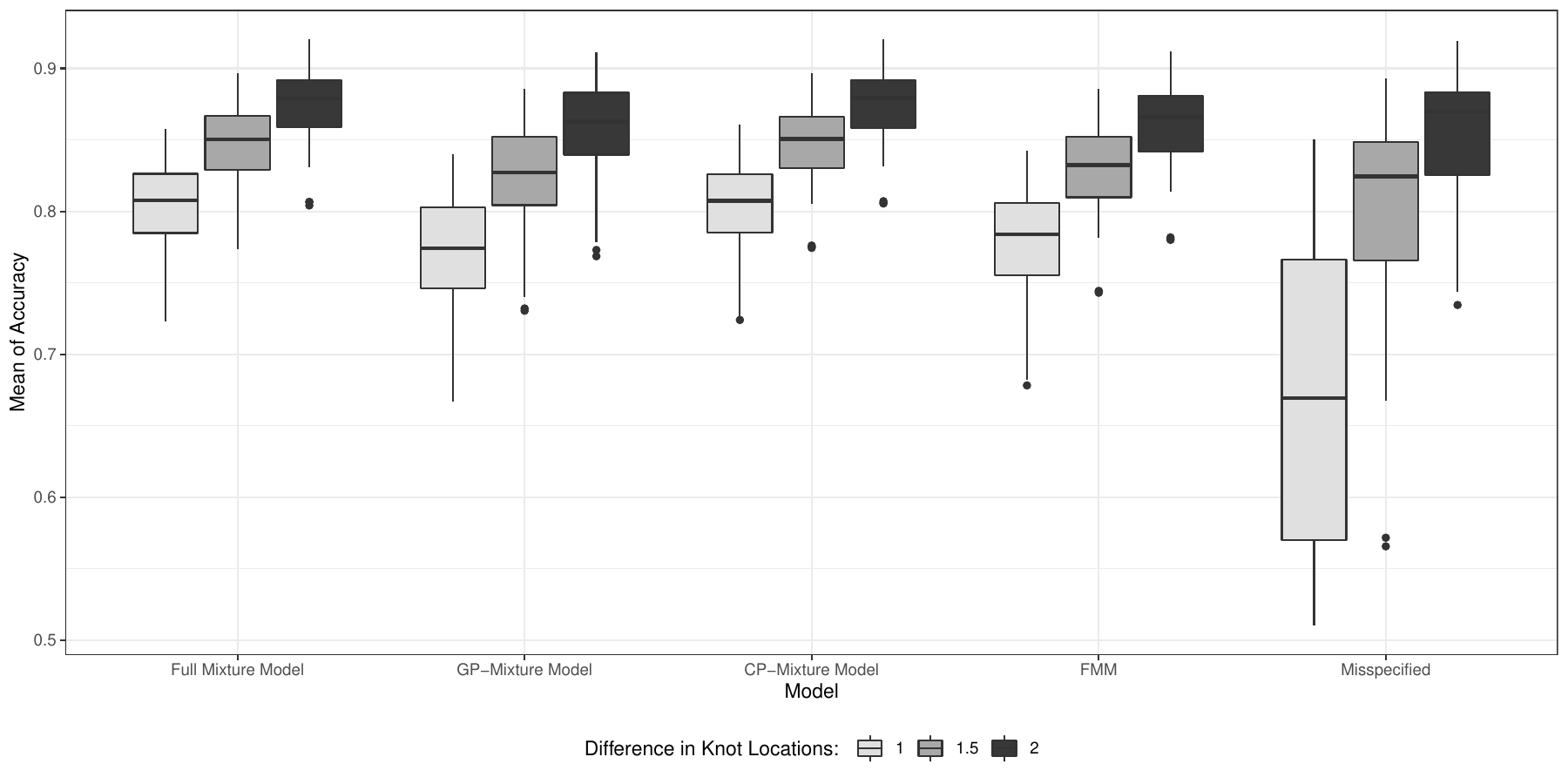}
\caption{Mean Values of Accuracy of Four Correctly-specified Models and One Misspecified Model}
\label{fig:Accuracy}
\end{figure}

\begin{figure}
\centering
\includegraphics[width=0.98\textwidth]{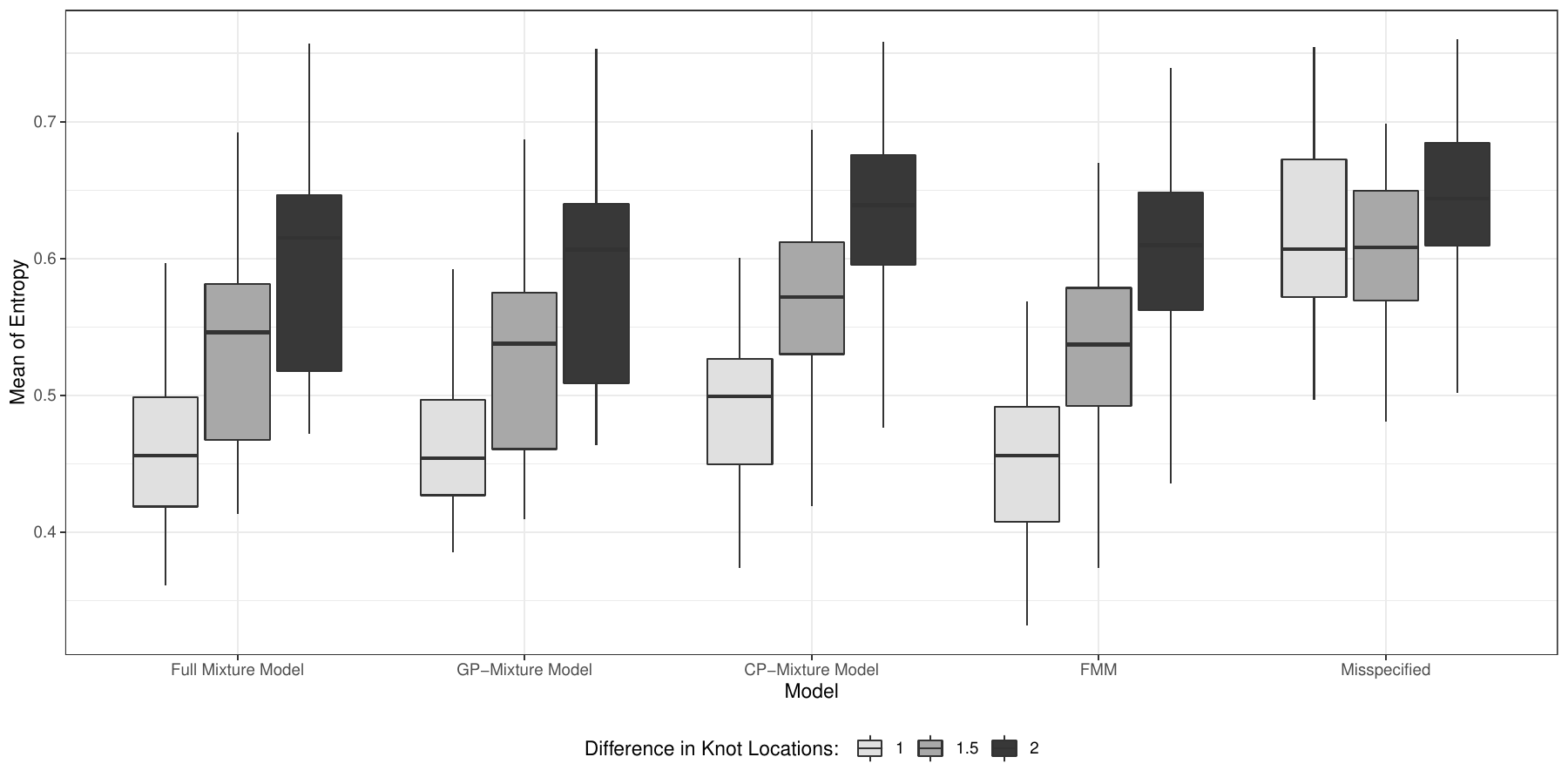}
\caption{Mean Values of Entropy of Four Correctly-specified Models and One Misspecified Model}
\label{fig:Entropy}
\end{figure}

\begin{figure}
\centering
\includegraphics[width=0.98\textwidth]{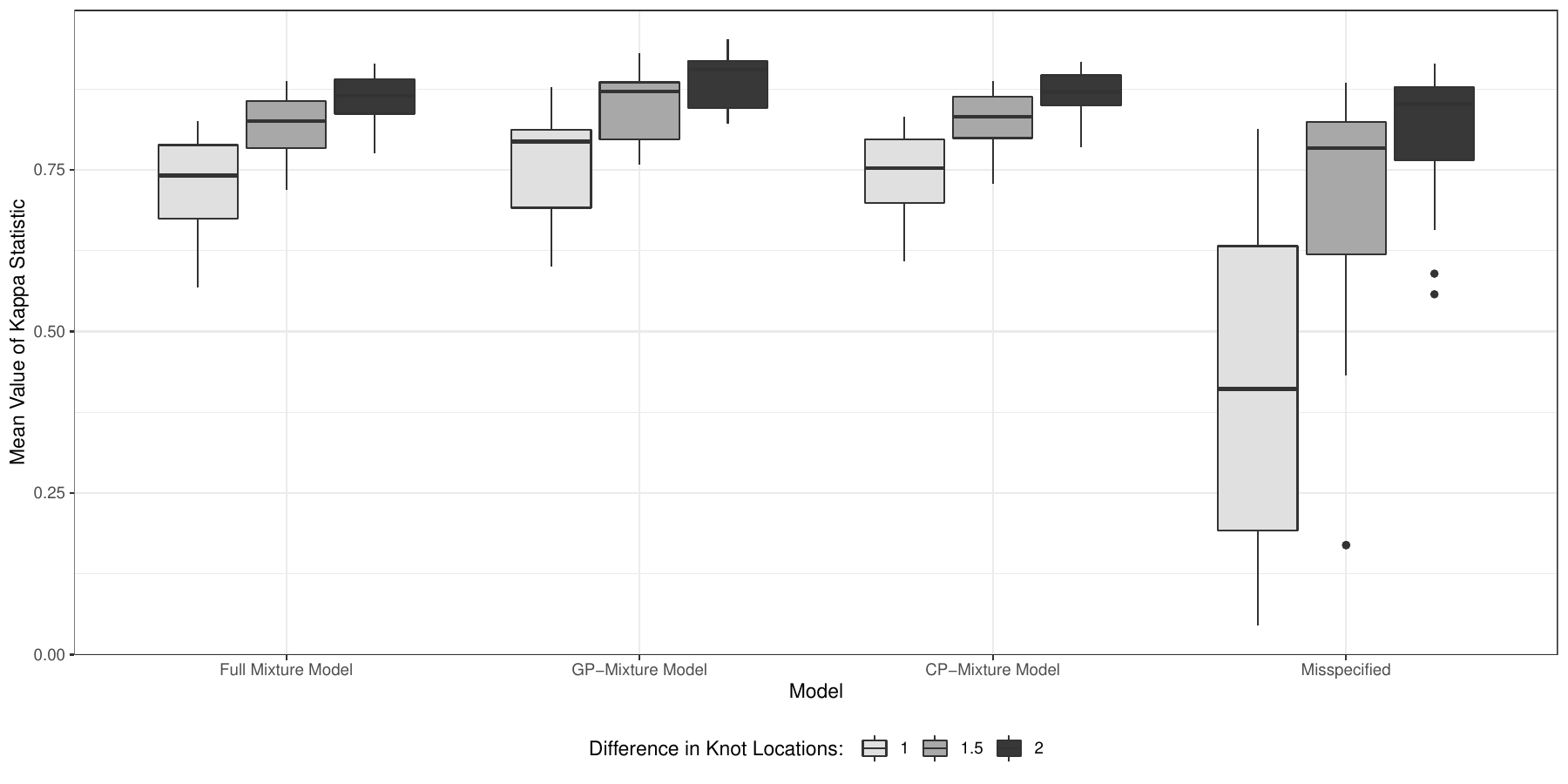}
\caption{Agreement between the Membership Obtained from  Finite Mixture Model and Each Model with Covariates}
\label{fig:Agree}
\end{figure}

\begin{figure}
\begin{subfigure}{.5\textwidth}
\centering
\includegraphics[width=0.9\linewidth]{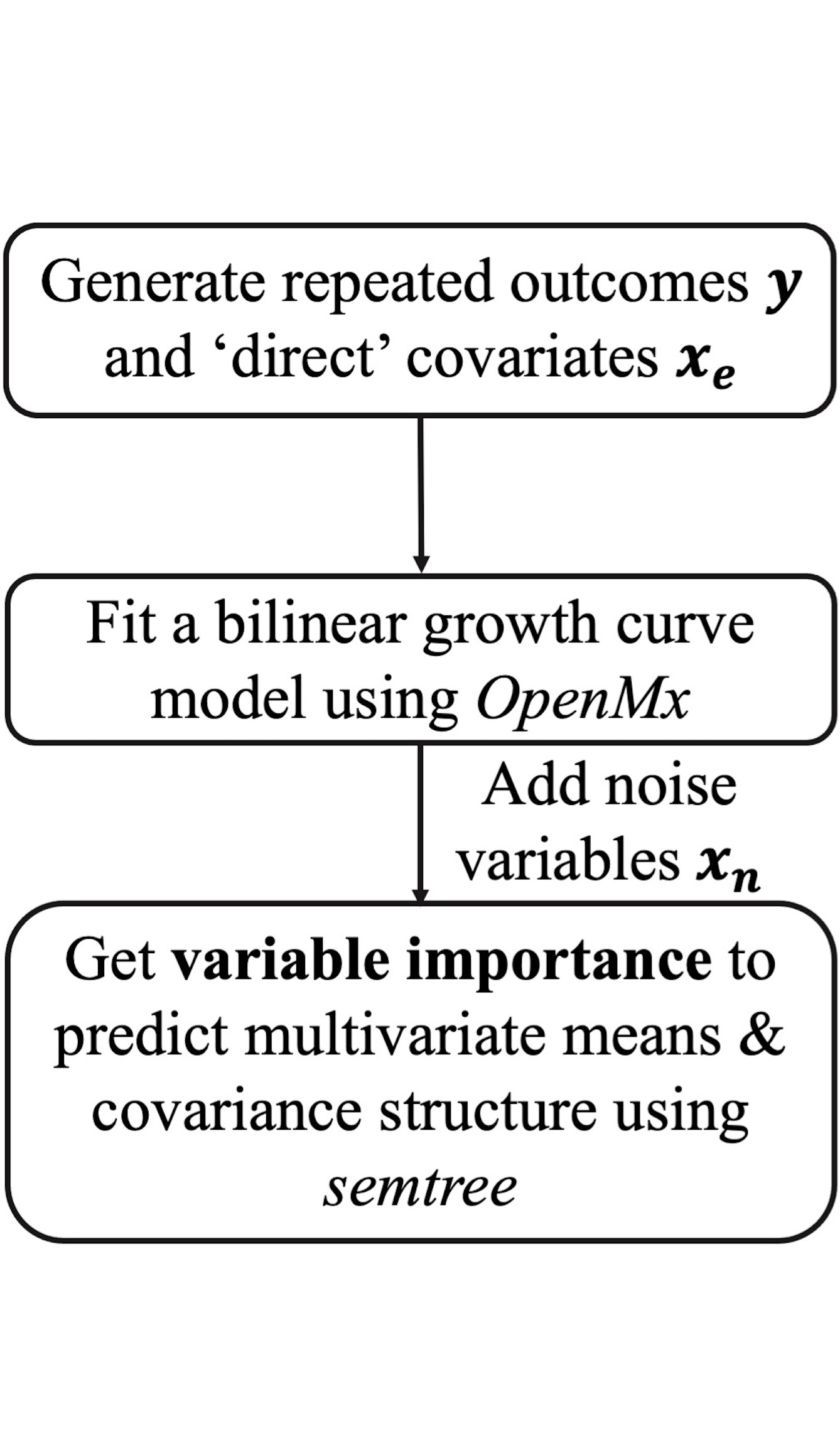}
\caption{One Class}
\label{fig:Step1}
\end{subfigure}%
\begin{subfigure}{.5\textwidth}
\centering
\includegraphics[width=0.9\linewidth]{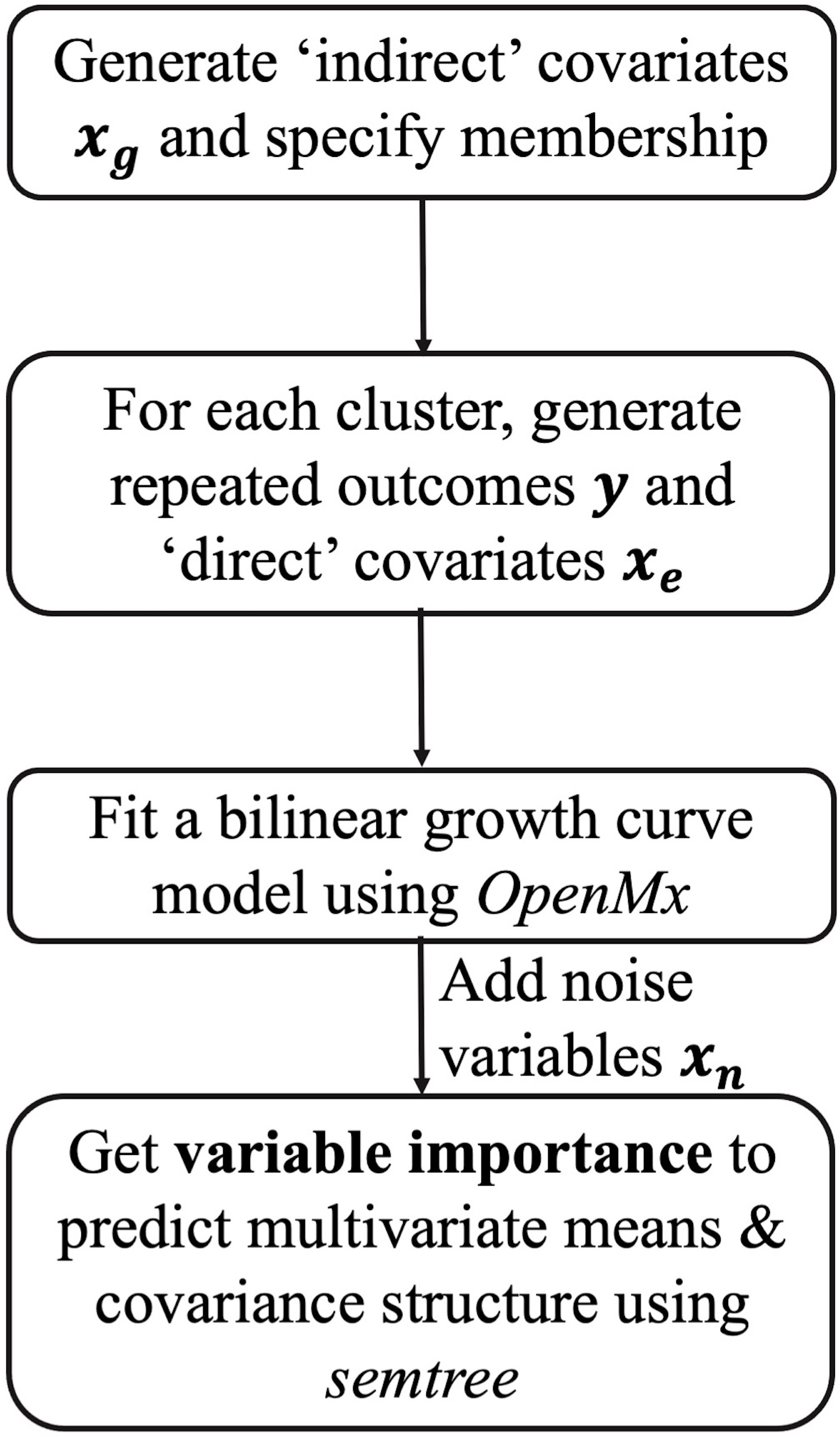}
\caption{Two Latent Classes}
\label{fig:Step2}
\end{subfigure}
\caption{General Steps to Obtain Covariates with High Importance for Heterogeneity of Nonlinear Trajectories}
\label{fig:Steps}
\end{figure}

\begin{figure}
\centering
\includegraphics[width=1.0\textwidth]{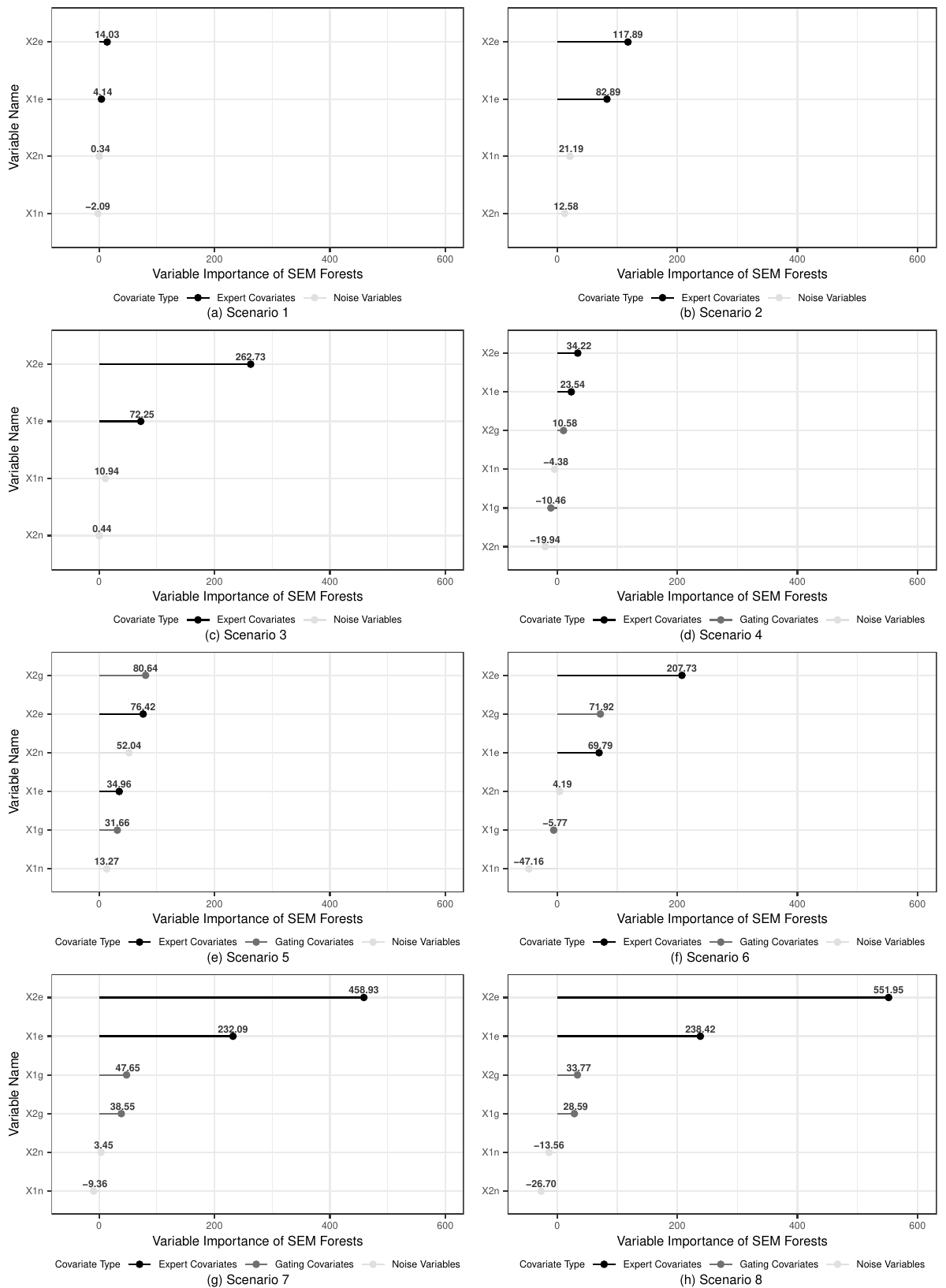}
\caption{Variable Importance Generated by SEM Forests for All Scenarios}
\label{fig:varImp}
\end{figure}

\begin{figure}
\centering
\includegraphics[width=1.0\textwidth]{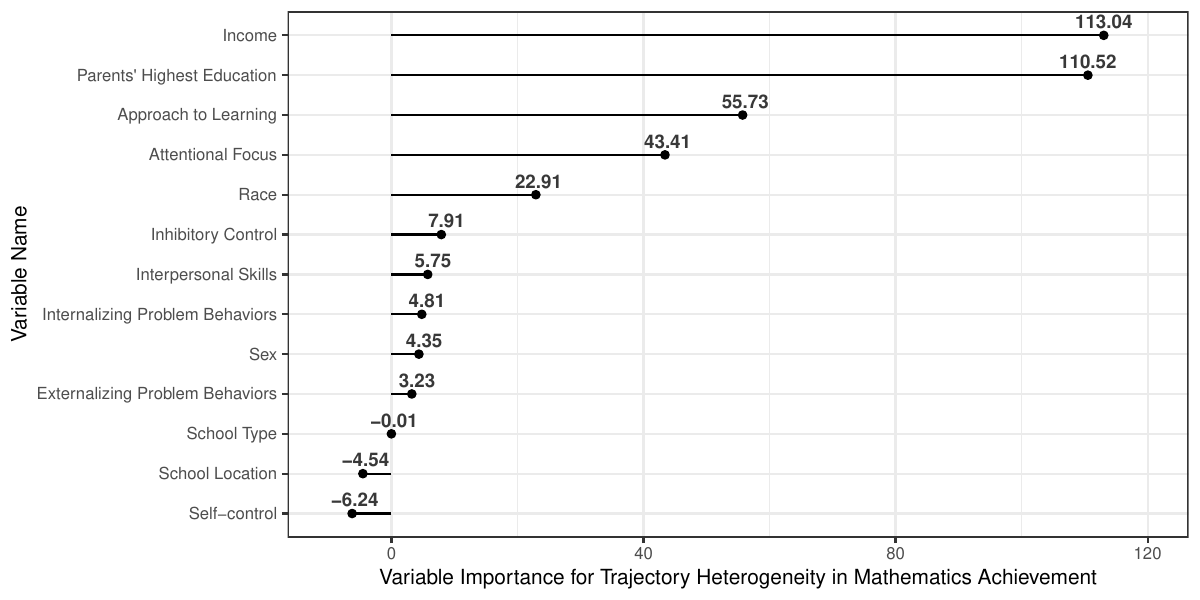}
\caption{Variable Importance Generated by SEM Forests for Nonlinear Trajectories of Mathematics Ability}
\label{fig:MathvarImp}
\end{figure}

\begin{figure}
\begin{subfigure}{.50\textwidth}
\centering
\includegraphics[width=1.0\linewidth]{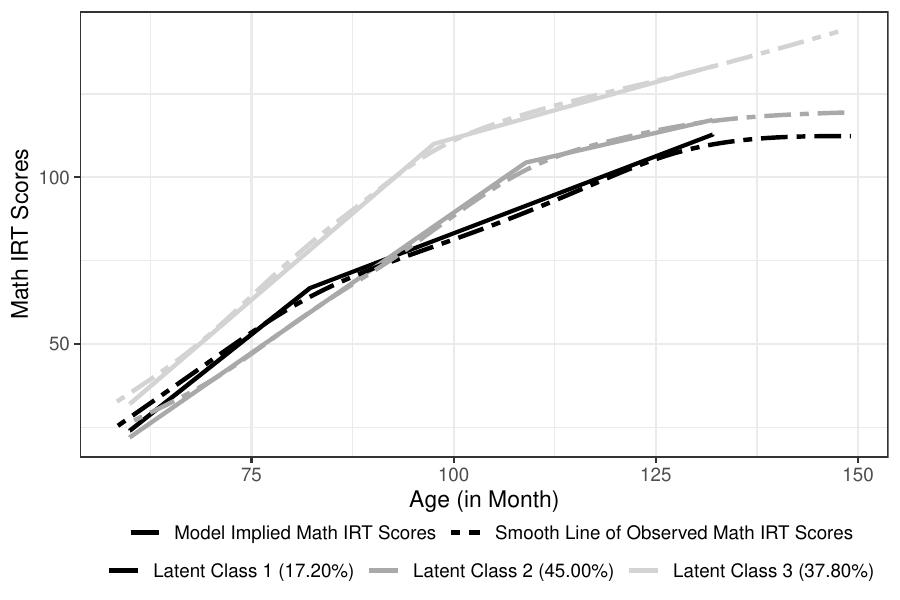}
\caption{Trajectories of Model 1}
\label{fig:FMM_traj}
\end{subfigure}%
\begin{subfigure}{.50\textwidth}
\centering
\includegraphics[width=1.0\linewidth]{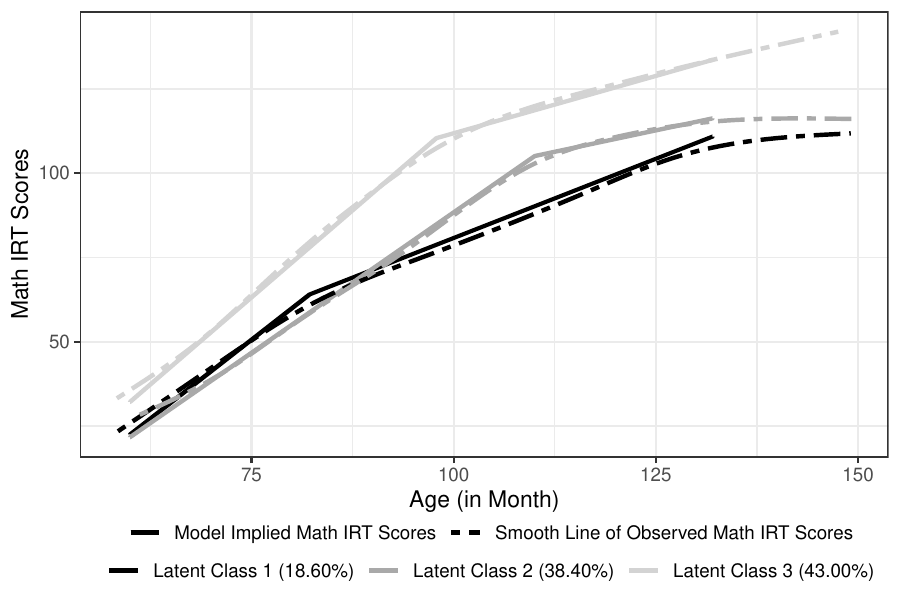}
\caption{Trajectories of Model 2}
\label{fig:Gating_traj}
\end{subfigure}
\vskip\baselineskip
\begin{subfigure}{.50\textwidth}
\centering
\includegraphics[width=1.0\linewidth]{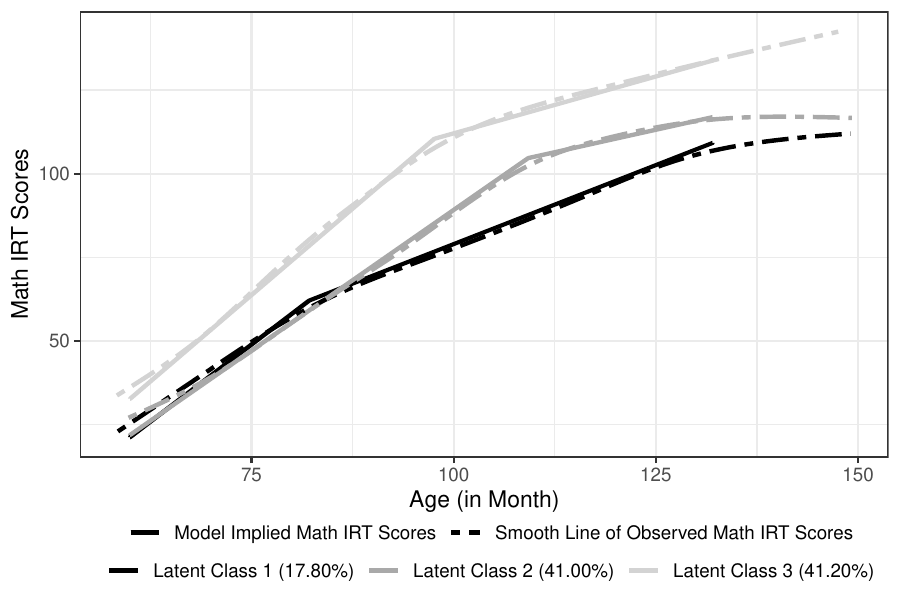}
\caption{Trajectories of Model 3}
\label{fig:misGating_traj}
\end{subfigure}%
\begin{subfigure}{.50\textwidth}
\centering
\includegraphics[width=1.0\linewidth]{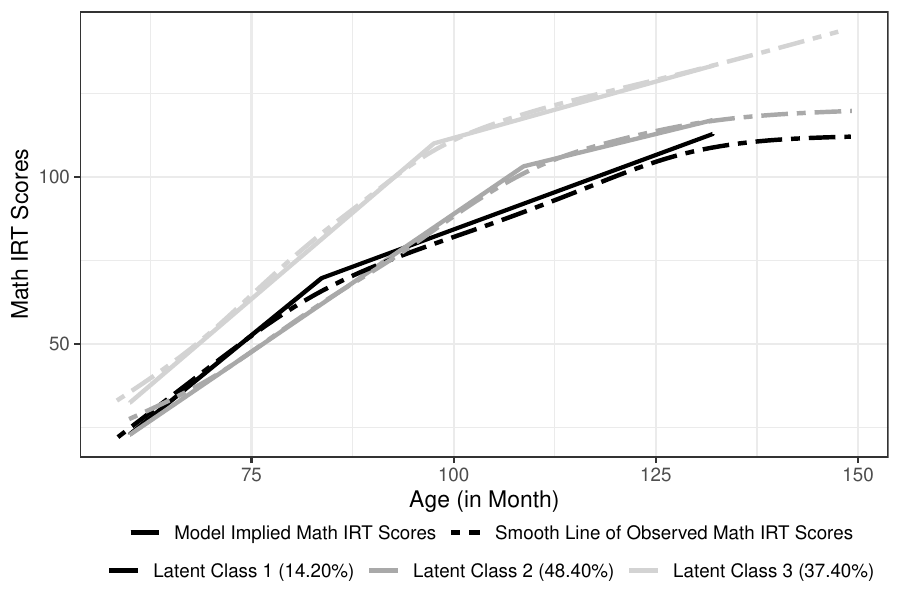}
\caption{Trajectories of Model 4}
\label{fig:Expert_traj}
\end{subfigure}
\vskip\baselineskip
\begin{subfigure}{.50\textwidth}
\centering
\includegraphics[width=1.0\linewidth]{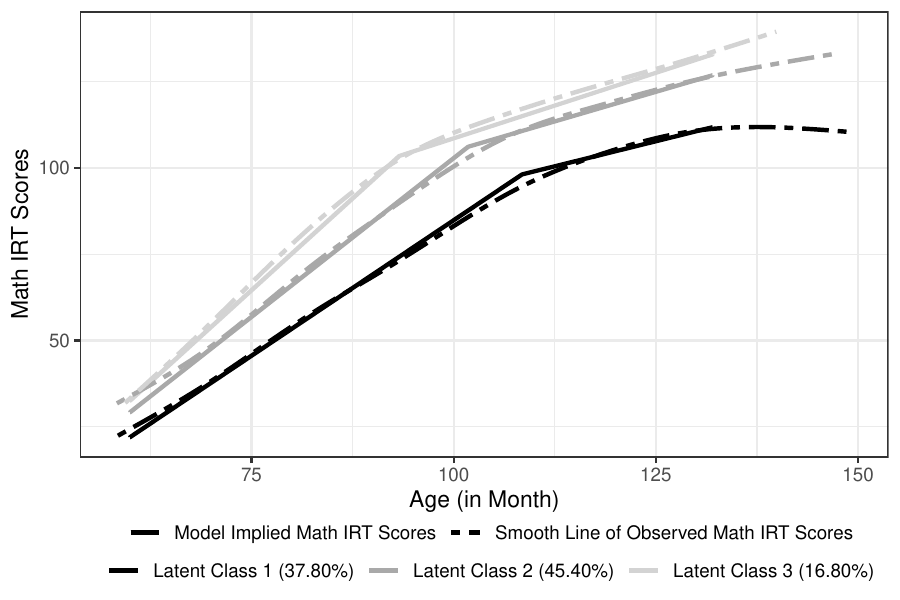}
\caption{Trajectories of Model 5}
\label{fig:misExpert_traj}
\end{subfigure}%
\begin{subfigure}{.50\textwidth}
\centering
\includegraphics[width=1.0\linewidth]{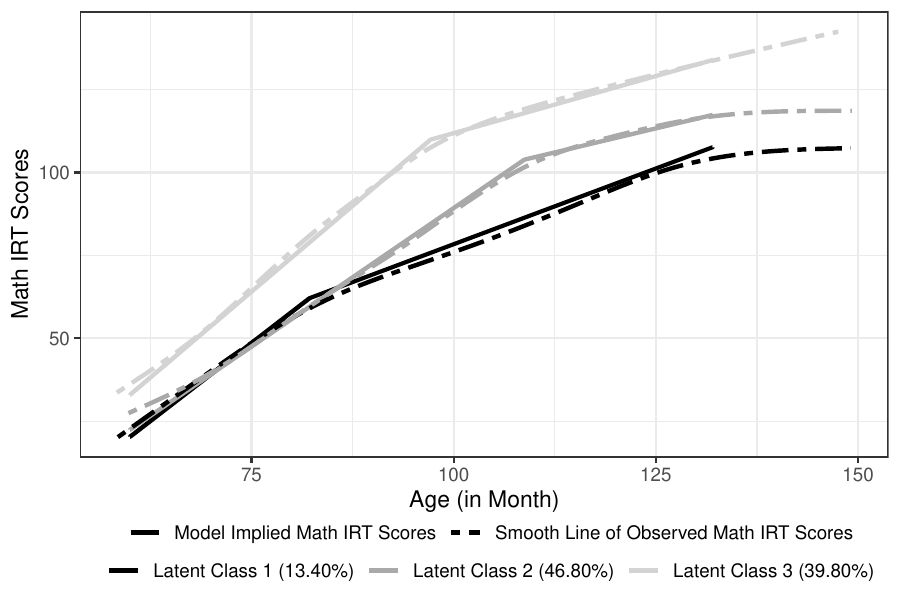}
\caption{Trajectories of Model 6}
\label{fig:Full_traj}
\end{subfigure}
\caption{Predicted Trajectories for Each of Three Latent Classes}
\label{fig:traj1}
\end{figure}

\renewcommand{\thefigure}{A.\arabic{figure}}
\setcounter{figure}{0}
\begin{figure}
\centering
\includegraphics[width=1.0\textwidth]{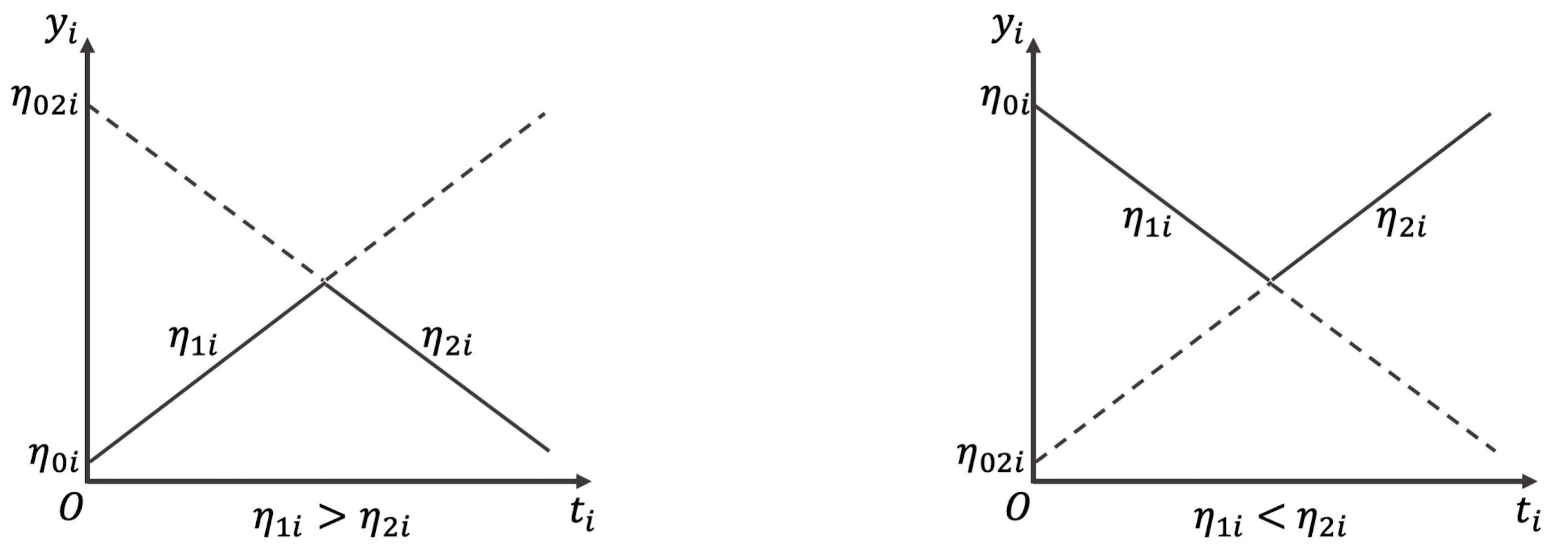}
\caption{The Two Forms of the Bilinear Spline (Linear-Linear Piecewise) }
\label{fig:E2_2cases}
\end{figure}

\end{document}